\documentclass[submit]{elsarticle}

\usepackage{hyperref}
\usepackage{multirow}
\usepackage{listings}
\usepackage{rotating}

\usepackage{soul} % for highlight
\usepackage[version=3]{mhchem} % for chemical formulas
\usepackage{multicol}
\usepackage{perpage}
\usepackage{siunitx}
\usepackage{amsmath}
\usepackage{longtable}
\usepackage{pdflscape} % for 'landscape' environment

\usepackage[version=3]{mhchem}
\usepackage{float}
\usepackage[table,xcdraw]{xcolor}
\journal{Journal of Planetary and Space Science}

\biboptions{authoryear}

\begin{document}

\begin{frontmatter}

%\tableofcontents
%\title{Mineralogical characterization using the online GIS client PlanetServer-2}
\title{Online characterization of planetary surfaces: PlanetServer, an open-source analysis and visualization tool.}
%%%%%%%%%%%%%%%%%%%%%%%%%%%
%% COMMENT i would try not to just limit the tile as a client, as also the server + python/api side of it are of interest even more to the community (remember the title/focus of the special issue...)

%\author{R. Marco Figuera\fnref{myfootnote}, B. Pham Huu, A. P. Rossi, M. Minnin}
%\address{Campus Ring 1, Bremen}

%\author{J. Flahaut}
%\address{rue de mart, lyon}

%\author[mymainaddress,mysecondaryaddress]{Jacobs University Bremen}

\author[]{R. Marco Figuera\corref{mycorrespondingauthor}}
\cortext[mycorrespondingauthor]{Corresponding author}
\ead{r.marcofiguera@jacobs-university.de}

\author[mymainaddress]{B. Pham Huu}
\author[mymainaddress]{A. P. Rossi}
\author[mymainaddress]{M. Minin}

\author[mysecondaryaddress]{J. Flahaut}
\author[mymainaddress]{A. Halder}

\address[mymainaddress]{Jacobs University Bremen, Campus Ring 1, 28759, Bremen, Germany}
\address[mysecondaryaddress]{Institut de Recherche en Astrophysique et Plan\'{e}tologie, UMR 5277 du CNRS, Universit\'{e} Paul Sabatier, 31400 Toulouse, France.}

\begin{abstract}
The lack of open-source tools for hyperspectral data visualization and analysis creates a demand for new tools. In this paper we present the new PlanetServer, a set of tools comprising a web Geographic Information System (GIS) and a recently developed Python Application Programming Interface (API) capable of visualizing and analyzing a wide variety of hyperspectral data from different planetary bodies. Current WebGIS open-source tools are evaluated in order to give an overview and contextualize how PlanetServer can help in this matters. The web client is thoroughly described as well as the datasets available in PlanetServer. Also, the Python API is described and exposed the reason of its development. Two different examples of mineral characterization of different hydrosilicates such as chlorites, prehnites and kaolinites in the Nili Fossae area on Mars are presented. As the obtained results show positive outcome in hyperspectral analysis and visualization compared to previous literature, we suggest using the  PlanetServer approach for such investigations. 
%%%%%%%%%%%%%%%%%%%%%%%%%%%
\end{abstract}


\end{frontmatter}

%\linenumbers

%\tableofcontents
\section{Introduction}
\subsection{Scientific Background}

The mineral characterization of planetary surfaces bears great importance for space exploration. Several studies have targeted a variety of minerals on Mars' surface based on imaging spectrometer data (\cite{Murchie2009}, \cite{Bibring2006}, \cite{Liu2016}, \cite{Cuadros2013}, \cite{Brown2010a}, \cite{Poulet2005} and \cite{Ehlmann2009}). Most of those studies have been carried out using VNIR spectroscopic data from the Observatoire pour la Min\'{e}ralogie, l'Eau, les Glaces et l'Activit\'{e} (OMEGA, Mars Express) (\cite{Bibring2004a}) and Compact Reconnaissance Imaging Spectrometer for Mars (CRISM, Mars Reconnaissance Orbiter)(\cite{Murchie2007}) imagery combined with different Digital Terrain Models (DTMs) of the areas. Accessing hyperspectral data stored in different planetary science data archives is a straightforward process. Nevertheless, analysis tools are often costly and methods tend to be hard to implement. 
%In order to enhance mineralogical signals in hyperspectral images, band math is widely used. \cite{Pelkey2007} and \cite{Viviano-Beck2014} developed CRISM specific band math combinations (CRISM summary products) which, used in a unique RGB combination, will highlight different minerals. 
Usually, the software used to process as well as to analyze data is the ENvironment for Visualizing Images (ENVI) package. Both OMEGA and CRISM experiment teams provided the science community with a set of IDL/ENVI routines (SOFT for OMEGA, CAT for CRISM) that allow basic processing steps. CAT for CRISM also includes the CRISM products described in \cite{Viviano-Beck2014} allowing to create RGB band math combinations. 
In this document we present the current status of PlanetServer – an open-source and ready-to-use online tool providing access to visualization and analysis of Mars CRISM and Moon Mineralogy Mapper (M3) images using the Web Coverage Processing Service (WCPS) Open Geospatial Consortium (OGC) standard, with the possibility of supporting more experiment and target Solar System bodies. 

\subsection{State of the art in planetary web services}

In the last years several planetary web services have been developed. Planetary web services cover a broad list of needs for the scientific community: from data exploration and retrieving to analysis and exploitation. Mars trek\footnote{http://marstrek.jpl.nasa.gov}, a web visualization service provided by NASA, showcases different data collected by a variety of instruments at selected landing sites. In this case, the service is more an outreach tool rather than an analysis tool. The service also offers data from the Moon\footnote{http://moontrek.jpl.nasa.gov} (\cite{Day2016a}) and Vesta\footnote{http://vestatrek.jpl.nasa.gov}. The PDS geoscience node also provides the so called Orbital Data Explorer (ODE)\footnote{http://ode.rsl.wustl.edu},a set of tools for visualization of Mars, Moon, Mercury and Venus data stored in PDS. The ODE Map search allows to focus on a Region of Interest (RoI) and select the datasets to download using point or polygon selection. A feature shared among the majority of visualization services is the selection of features by clicking or by a polygon and the change of cartographic projection.

The recently developed Multi-Temporal Database of Planetary Image Data (MUTED) (\cite{Heyer2016}) is an online web service providing different spatial and temporal Martian datasets based on a polygon selection\footnote{http://muted.wwu.de/}. An asset of MUTED is the possibility to retrieve data based on temporal or spatial parameters of a selected RoI, thus making it very convenient and appropriate when multiple multi-temporal datasets exist. Another web service is i-mars \footnote{http://www.i-mars.eu/web-gis} (\cite{VanGasselt2014})providing online visualization of selected datasets.

While the above mentioned services focus on the visualization and data retrieval, different services exist that allow the user to process and analyze data online. The MARS Information System (MarsSI)\footnote{https://emars.univ-lyon1.fr/MarsSI} (\cite{Lozach2015}),a web application developed at the Universit\'{e} Lyon 1, allows the user to select raw Martian data (HiRISE, CTX, CRISM, OMEGA, etc) and process them using built-in pipelines. The output is data that can be loaded into a GIS. While MarsSI allows selection and processing of data it does not visualize the final product in-situ neither allows to interactively analyze the data. In order to use MarsSI one needs to register by filling a form provided at their website, thus making the access to the tool cumbersome. The Planetary SUrface Portal (PSUP)\footnote{http://psup.ias.u-psud.fr} (Poulet et al., this special issue) combines in one platform the processing power of MarsSI together with the visualization tool MarsVisu. PSUP allows the user to analyze and visualize data from a wide variety of datasets within the same environment.

The old version of PlanetServer (\cite{Oosthoek2014a} and \cite{Oosthoek2015}) developed within the EU FP7-INFRA project EarthServer was the predecessor and initial client of the actual PlanetServer. The old client provided visualization and analysis of CRISM data through a web portal using openlayers and self made tools. The visualization was presented as a 2D map and the analysis and selection tools were displayed on top of the main map. The architecture of the old PlanetServer was very  similar to the actual client, using rasdaman as the database manager software and WCPS queries in order to access the data.

If we analyze the capabilities of the above mentioned web services, with the exception of the old PlanetServer, we quickly realize the lack of in-situ analysis tools and online output files visualization. With the new version of PlanetServer we aim to cover this gap in the current available web services by adding the possibility to analyze data in real-time and present the output results without leaving the web environment. This is possible by adopting the WCPS OGC standard discussed in detail in the following chapters.

After a description of the architecture, datasets and features of PlanetServer (section \ref{sec:planetserver}), we introduce the PlanetServer Python API (section \ref{sec:api}). Afterwards, we demonstrate the tool reliability with some examples of analysis carried out with PlanetServer (section \ref{sec:examples}). Finally, discussions and ongoing work are presented.

\section{PlanetServer}
\label{sec:planetserver}

PlanetServer is the planetary service of the H2020 EC-funded project EarthServer-2 (\cite{Baumann2015}) aiming at access and exploitation of Big Earth Science data. The project comprises various services related to specific domains (Earth Observation, Marine Remote Sensing and Climate Modeling). The planetary section of the project is developed at Jacobs University Bremen and it focuses on the visualization and analysis of space mission data on solid planets and moons, using OGC standards on a web client based on the JavaScript version of NASA's World Wind (\cite{Hogan}) and a Python API. PlanetServer is currently divided in two web clients containing Mars CRISM data and Moon Mineral Mapper (M3) (\cite{Pieters2009}, \cite{Green2011}) data and a Python API capable of accessing and analyzing both datasets. The Python API will be further discussed in section \ref{sec:api}. 

\subsection{Architecture}
\label{sec:architecture}

\begin{figure}[H]
\centering
\includegraphics[width=300pt]{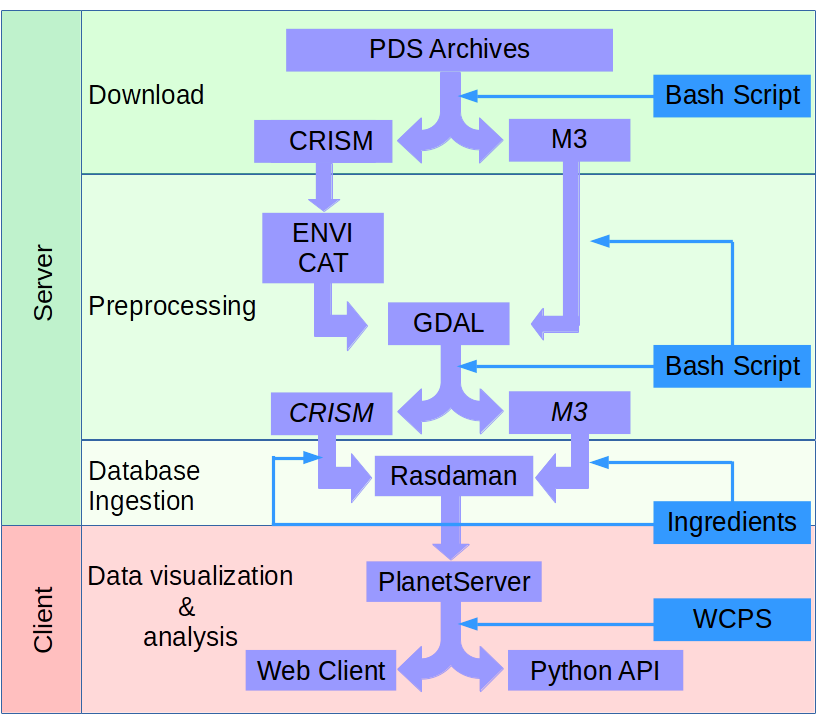}%ext=bmp,jpg,...
\caption[PlanetServer pipeline.]{This figure shows the PlanetServer pipeline covering all the process. The data is downloaded from PDS archive, processed using ENVI CAT and GDAL for CRISM, and only GDAL for M3. Data are ingested into Rasdaman. PlanetServer's data can be accessed using the web client or the Python API.  }
\label{fig:diagram}
\end{figure}

As Figure \ref{fig:diagram} highlights, the service architecture has two components: the server and the client side. In the server side the download, preprocessing and storage of the data takes place.

M3 and CRISM datasets are publicly available in the Planetary Data Service (PDS) archives (\cite{McMahon1996}), after limited embargo periods. Therefore the retrieval of planetary data from different missions and instruments is a straightforward process. The data are downloaded using a set of dedicated scripts that search for the entire CRISM and M3 catalogs in the PDS archive. Data reduction and processing historically relies largely on users, although recent initiatives provide server side pre-processing (\cite{Hare}). The data is subsequently processed in our servers in order to apply the necessary atmospheric corrections and map projections. A more detailed explanation of the data download and preprocessing is explained in section \ref{sec:process-data}. Finally, data are stored using the Array DataBase Management System (DBMS) Raster Data Manager (Rasdaman), capable of storage and retrieval of multidimensional data (\cite{Baumann1998}). Rasdaman is an Array DBMS offering features such as query languages, query optimization and parallelization on n-D arrays. The Rasdaman data model consists of n-D arrays with individually fixed or variable boundaries. The query language, ”rasql”, is an ISO SQL extended language allowing declarative array selection and processing with multidimensional operators. The array is partitioned (tiled) and stored along with the processing engine, thus reducing the query response time. OGC standards such as the WCPS (\cite{Baumann2009a}), are implemented in the PetaScope component (\cite{Aiordachioaie2010}), a set of geospatial and geometry libraries, data access libraries and relational database access components.

In the client side, two different methods are available for data visualization and analysis. Using the web client, the data is visualized using NASA’s Web World Wind allowing access through a web interface. NASA Web World Wind is a general-purpose 3D/4D client used as a virtual globe to interactively analyze and visualize data. As Web World Wind is completely open-source, one can easily extend its functionality and Application Programming Interface (API) to fit in the project purposes. The API has been developed for HTML5 and JavaScript and uses WebGL as a rendering engine. PlanetServer extensively uses OGC standards (WMS and WCPS) in order to retrieve, visualize and compute data. The web client contains two main features, the RGB combinations and Spectral analysis tools. Using the Python API or Jupyter notebooks the user has access to the same features as in the web client. Data visualization and analysis using PlanetServer's Python API is further discussed in section \ref{sec:api}.
 
\subsubsection{Python API}
\label{sec:api}
As Python is gaining a remarkable popularity among the planetary science and astronomy community, we developed the PlanetServer Python API. The tool is available as a Python package that can be integrated in existing python pipelines as well as a Jupyter notebook, capable of running in a browser. The API integrates the features available in the web client: RGB combinations and spectral analysis tools. 

In order to create the CRISM products, we created a dictionary in JSON format where a set of parameters are given:

\begin{lstlisting}[ language=SQL,
           caption={Declaration of CRISM products in JSON format.},
           label=lst:crism,
           showspaces=false,
           basicstyle=\footnotesize,
           breaklines=true,
           numbers=left,
           numberstyle=\tiny,
           commentstyle=\color{gray}
        ]
summary_products = { 
"BD1300" : [f1, ["R1320", "R1080", "R1750"]], 
"BD1400" : [f1, ["R1395", "R1330", "R1467"]], 
"BD1435" : [f1, ["R1435", "R1370", "R1470"]], 
"BD1500_2" : [f1, ["R1525", "R1367", "R1808"]], 
...
} 
\end{lstlisting}

The first parameter (BD1300) is the name of the CRISM product. This name will we used when creating the RGB combination. The second parameter (f1) is the family type. CRISM products have been grouped based on their similarities\footnote{The family definition is available on the Jupyter notebook inside PlanetServers' GitHub repository: https://github.com/planetserver/PS\_Python\_API} thus speeding the definition process. The last parameters are the wavelengths that compose the product ($R_c$ being the central wavelength, $R_s$ the shorter wavelength and $R_L$ the longer wavelength, given in this order). 

\begin{figure}[H]
\centering
\includegraphics[width=400pt]{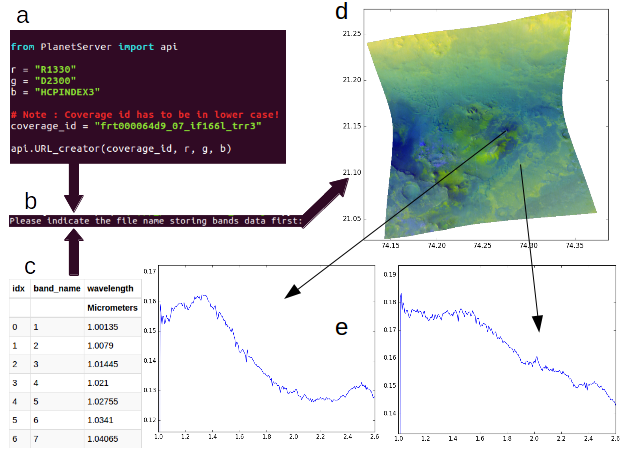}%ext=bmp,jpg,...
\caption{PlanetServer Python API workflow. a) The user inputs the three CRISM summary products in the corresponding RGB channels (R:LCPINDEX2, G:D2200 and B:HCPINDEX3) and the coverage ID. b) A Flexible Image Transport System (FITS) table (provided in the package) with the relation between band names and wavelengths (c) needs to be called in order to create correct WCPS queries. d) The API outputs an image containing the RGB combination. e) By clicking a location in the image (e.g: 21.19$^{\circ}$N, 74.24$^{\circ}$E), the API will show the spectral profile.}
\label{fig:python_api}
\end{figure}

As illustrated in Figure\ref{fig:python_api}, the Python API has been constructed such that a user only needs to give five input arguments to the API: The coverage ID, the name of three CRISM summary products that are stored in each channel (Red, Green and Blue) and the name of a table containing the relation between the band names and the corresponding wavelengths. The table has been created using Astropy and extracting the metadata attached to any CRISM image. The script to create tables for new datasets is provided in the package. The API combines the 3 summary products into a WCPS query giving an image as output. The spectral analysis tool will be activated by clicking a point in the image. All the plots and images can be downloaded for further editing.

\subsection{Features}

\begin{figure}[H]
\centering
\includegraphics[width=400pt]{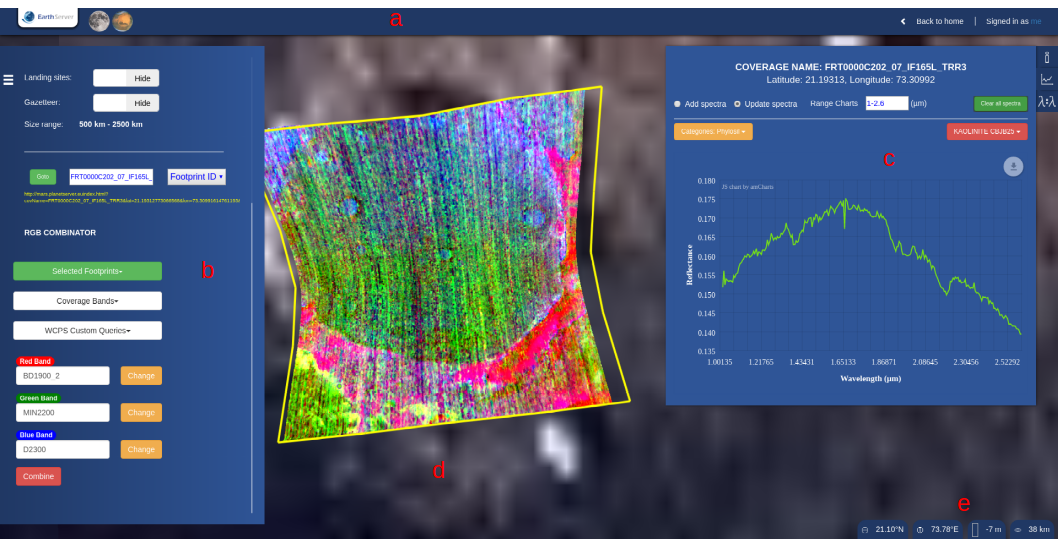}%ext=bmp,jpg,...
\caption[PlanetServer outlook.]{PlanetServer web client outlook. a) The navigation bar is shown in the upper part of the client. It allows the user to navigate through the different available planetary bodies. b) The main dock, located in the left part, contains the layers, search and RGB combination tool. c) shows the plotting dock where all the spectral analysis is carried out. d) The main globe accommodates the georeferenced analyzed images. e) The latitude, longitude, altitude as well as height information is shown.}
\label{fig:global}
\end{figure}

PlanetServer's web client has been developed in a non-invasive and minimalist way. The client is divided in 4 areas (Figure \ref{fig:global}) containing the Globe and different docks.

\begin{itemize}
\item Navigation bar: The navigation bar (located in the upper part) contains links to the service description website and the project description website as well as the Moon and Mars icon to navigate between planets.
\item Main dock: Located in the left part of the client, the main dock contains: projections, available base maps, search engine (by location, name and coordinates) and the RGB combination tool (this feature is further discussed in section \ref{sec:rgb}).
\item Secondary dock: Located in the right part of the client, the secondary dock contains 3 sub-docks: The spectral plot and spectral ratio dock (where the spectral analysis tool (discussed in section \ref{sec:spectral}) is used), and the information dock (which contains information about the client, a contact form as well as a tour, guiding the user through the client showing all the functionalities).
\item Globe: In the central part of the client the Globe is shown. Depending on the client, either the Moon or Mars will be loaded.
\item Coordinate and Elevation: Located in the bottom right part of the client are the coordinates, elevation and eye altitude information at the current position.
\end{itemize}

PlanetServer uses extensively OGC standards in order to retrieve graphic information. WMS is an OGC standard allowing the user to retrieve images hosted on a server. In PlanetServer, WMS is used to retrieve base maps as well as deploy the DTMs. An innovative feature of PlanetServer is the use of WCPS, allowing the user to retrieve images by setting different parameters such as area, time, bands, etc. The main difference to other OGC standards is that the WCPS retrieved image is not static and can be changed by updating the parameters.

\subsubsection{RGB combination tool}
\label{sec:rgb}

The RGB combination tool allows the user to create different RGB combinations by selecting single or multiple images and a set of bands or CRISM products to be combined in each channel. As multiple selection is possible, the same RGB combination can be processed in several images at once. Also, when images overlap the user can select the desired image and lock it, thus targeting the desired image.

One of the key features of PlanetServer is the possibility to retrieve different CRISM summary products allowing to pursue better enhance and better visualize different materials in an image. We have translated the CRISM summary products (\cite{Viviano-Beck2014}) to WCPS and allow access through the web client and the Python API. A list of 44 summary products are currently available (Table \ref{tab:products}). The tool combines three CRISM products in each RGB channel and creates a WCPS query to compute the combination. The output of this query is a TIFF image which is temporary downloaded into our servers. In order to enhance and better visualize the image we set a fixed stretching values. The mean value ($\mu$) of each channel as well as the standard deviation ($\sigma$) is calculated using the Geospatial Data Abstraction Library (GDAL) and the image is re-stretched using as minimum value $\mu - 1.5 * \sigma$ and maximum value $\mu + 1.5 * \sigma$. The image is then send back to PlanetServer in PNG format to be displayed in the globe. Currently, this stretching values are fixed and shared among all the RGB combinations. Ongoing work includes the possibility to dynamically change the stretching values according to the user needs. At present, the CRISM RGB combinations created with PlanetServer are not considering any thresholds and all data is taken into account. Ongoing work includes setting a threshold greater than 0 in order to avoid the background noise. As stated by \cite{Carter2013a} several artifacts can occur in the RGB combinations when specific corrections are not applied to the original data.  We are currently studying the possibility of adding such corrections in order to prevent such artifacts.

\begin{figure}[H]
\centering
\includegraphics[width=\textwidth]{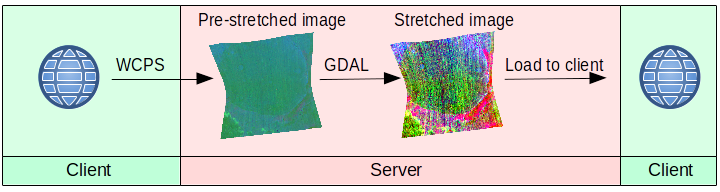}%ext=bmp,jpg,...
\caption{Worflow of the RGB combination. The tool combines the selected CRISM products and creates a WCPS query. The query computes an image where the minimum and maximum values cover the entire histogram. GDAL is run on the image to extract the mean value and standard deviation of each band. After applying the new stretching values the image is then send to the client or Python API.}
\label{fig:rgb_comb}
\end{figure}

\begin{landscape}
\begin{longtable}{p{2cm} p{4cm} p{10cm} p{4cm}}
%\begin{tabular}{|p p{5cm} p{5cm}|}
\hline
\rowcolor[HTML]{C0C0C0}
Name & Parameter & Equation & Notes  \\ \hline \endhead
BD1300 & \SI{1.3}{\micro\metre} absorption associated with \ce{Fe^2+} & $1 - \left( \frac{R1320}{a \cdot R1080 + b \cdot R1750} \right)$ & Plaglioclase with \ce{Fe^2+} substitution  \\ \hline
BD1400 & \SI{1.4}{\micro\metre} \ce{H2O} and \ce{OH} band depth & $1 - \left( \frac{R1395}{a \cdot R1330 + b \cdot R1467} \right)$ & Hydrated or hydroxylated minerals  \\ \hline
BD1435 & \SI{1.435}{\micro\metre} \ce{CO2} ice band depth & $1 - \left( \frac{R1435}{a \cdot R1370 + b \cdot R1470} \right)$ & \ce{CO2} ice, some hydrated minerals  \\ \hline
BD1500\_2 & \SI{1.5}{\micro\metre} \ce{H2O} ice band depth & $1 - \left( \frac{R1525}{a \cdot R1367 + b \cdot R1808} \right)$ & \ce{H2O} ice on surface or in atmosphere   \\ \hline
BD1750\_2 & \SI{1.7}{\micro\metre} \ce{H2O} band depth & $1 - \left( \frac{R1750}{a \cdot R1690 + b \cdot R1815} \right)$ &  Gypsum, Alunite \\ \hline
BD2100\_2 & \SI{2.1}{\micro\metre} shifted \ce{H2O} band depth & $1 - \left( \frac{R2132}{a \cdot R1930 + b \cdot R2250} \right)$ &  \ce{H2O} monohydrated sulfates \\ \hline
BD2165 & \SI{2.165}{\micro\metre} \ce{Al-OH} band depth & $1 - \left( \frac{R2165}{a \cdot R2120 + b \cdot R2230} \right)$ & Pyrophyllite, kaolinites   \\ \hline
BD2190 & \SI{2.190}{\micro\metre} \ce{Al-OH} band depth & $1 - \left( \frac{R2185}{a \cdot R2120 + b \cdot R2250} \right)$ & Beidellite, Allophane, Imogolite  \\ \hline
BD2210\_2 & \SI{2.215}{\micro\metre} \ce{Al-OH} band depth  & $1 - \left( \frac{R2210}{a \cdot R2165 + b \cdot R2250} \right)$ & \ce{Al-OH} minerals  \\ \hline
BD2230 & \SI{2.23}{\micro\metre} band depth  & $1 - \left( \frac{R2235}{a \cdot R2210 + b \cdot R2252} \right)$ & Hydroxylated ferric sulfate   \\ \hline
BD2250 & \SI{2.25}{\micro\metre} broad \ce{Al-OH} and \ce{Si-OH} band depth  & $1 - \left( \frac{R2245}{a \cdot R2120 + b \cdot R2340} \right)$ & Opeal and other \ce{Al-OH} minerals   \\ \hline
BD2265 & \SI{2.265}{\micro\metre} band depth  & $1 - \left( \frac{R2265}{a \cdot R2210 + b \cdot R2340} \right)$ & Jarosite, Gibbsite, Acid-leached nontronite  \\ \hline
BD2290 & \SI{2.3}{\micro\metre} \ce{Mg}, \ce{Fe-OH} band depth and \SI{2.292}{\micro\metre} \ce{CO2} ice band depth & $1 - \left( \frac{R2290}{a \cdot R2250 + b \cdot R2350} \right)$ & \ce{Mg}, \ce{Fe-OH} minerals and \ce{CO2} ice  \\ \hline
BD2355 & \SI{2.53}{\micro\metre} band depth & $1 - \left( \frac{R2355}{a \cdot R2300 + b \cdot R2450} \right)$ & Chlorite, Prehnite, Pumpellyite  \\ \hline
BD2500\_2 & \ce{Mg} carbonate overtone band depth & $1 - \left( \frac{R2480}{a \cdot R2364 + b \cdot R2570} \right)$ & \ce{Mg} carbonates  \\ \hline
BD3100 & \SI{3.1}{\micro\metre} \ce{H2O} ice band depth & $1 - \left( \frac{R3120}{a \cdot R3000 + b \cdot R3250} \right)$ & \ce{H2O} ice  \\ \hline
BD3200 & \SI{3.2}{\micro\metre} \ce{CO2} ice band depth & $1 - \left( \frac{R3320}{a \cdot R3250 + b \cdot R3390} \right)$ & \ce{CO2} ice  \\ \hline
BD3400\_2 & \SI{3.4}{\micro\metre} carbonate band depth & $1 - \left( \frac{R3420}{a \cdot R3250 + b \cdot R3630} \right)$ & Carbonates  \\ \hline
BD2600 & \SI{2.6}{\micro\metre} \ce{H2O} band depth & $1 - \left( \frac{R2600}{a \cdot R2530 + b \cdot R2630} \right)$ & \ce{H2O} vapor  \\ \hline
MIN2200 & \SI{2.16}{\micro\metre} \ce{Si-OH} band depth and \SI{2.21}{\micro\metre} \ce{H-}bounds \ce{Si-OH} band depth & $ minimum \left[\left( 1 - \left( \frac{R2165}{a \cdot R2120 + b \cdot R2350} \right)\right) , \left( 1 - \left( \frac{R2210}{a \cdot R2120 + b \cdot R2350} \right)\right)\right]$ & Kaolinites  \\ \hline
MIN2250 & \SI{2.21}{\micro\metre} \ce{Si-OH} band depth and \SI{2.26}{\micro\metre} \ce{H-}bounds \ce{Si-OH} band depth & $ minimum \left[\left( 1 - \left( \frac{R2210}{a \cdot R2165 + b \cdot R2350} \right)\right) , \left( 1 - \left( \frac{R2265}{a \cdot R2165 + b \cdot R2350} \right)\right)\right]$ & Opals  \\ \hline
MIN2295\_2480 & \ce{Mg} carbonate overtone band depth and metal\ce{OH} band & $ minimum \left[\left( 1 - \left( \frac{R2295}{a \cdot R2165 + b \cdot R2364} \right)\right) , \left( 1 - \left( \frac{R2480}{a \cdot R2364 + b \cdot R2570} \right)\right)\right]$ & \ce{Mg} carbonates; both overtones must be present  \\ \hline
MIN2345\_2537 & \ce{Ca/Fe} carbonate overtone band depth and metal\ce{OH} band & $ minimum \left[\left( 1 - \left( \frac{R2345}{a \cdot R2250 + b \cdot R2430} \right)\right) , \left( 1 - \left( \frac{R2537}{a \cdot R2430 + b \cdot R2602} \right)\right)\right]$ & \ce{Ca/Fe} carbonates; both overtones must be present   \\ \hline
BD1900\_2 & \SI{1.9}{\micro\metre} \ce{H2O} band depth & $ 0.5 \cdot \left( 1 - \left( \frac{R1930}{a \cdot R1850 + b \cdot R2067} \right)\right) + 0.5 \cdot \left( 1 - \left( \frac{R1985}{a \cdot R1850 + b \cdot R2067} \right)\right)$& Bound molecular \ce{H2O} except monohydrated sulfates  \\ \hline
R1330 & IR albedo & $R1330$ & IR albedo (ices $>$ dust $>$ unaltered mafics)  \\ \hline
R1080 & \SI{1.08}{\micro\metre} reflectance & $R1080$ &  Component of the false color RGB combination \\ \hline
R1506 & \SI{1.51}{\micro\metre} reflectance & $R1506$ & Component of the false color RGB combination \\ \hline
R2529 & \SI{2.53}{\micro\metre} reflectance & $R2529$ & Component of the visible to infrared false color RGB combination  \\ \hline
R3920 & \SI{2.53}{\micro\metre} reflectance & $R3920$ &  Component used in the ice detection RGB combination\\ \hline
IRR2 & IR ratio 2 & $\frac{R2530}{R2210}$ & Aphelion ice clouds versus seasonal or dust  \\ \hline
IRR3 &  IR ratio 3 & $\frac{R3500}{R3390}$ &  Aphelion ice clouds (higher values) versus seasonal or dust  \\ \hline
BD3000 & \SI{3}{\micro\metre} \ce{H2O} band depth &  $1 - \left( \frac{R3000}{R2530 \cdot \left( \frac{R2530}{R2210} \right)} \right)$ &  Bound \ce{H2O} ( accounts for spectral slope)  \\ \hline
SINDEX2 & Inverse lever rule to detect convexity at \SI{2.29}{\micro\metre} due to \SI{2.1}{\micro\metre} and \SI{2.4}{\micro\metre} absorptions &  $1 - \left( \frac{a \cdot R2120 + b \cdot R2400}{R2290} \right)$ & Hydrated sulfates (mono and poly hydrated sulfates) will be strongly $>$ 0  \\ \hline
CINDEX2 & Inverse lever rule to detect convexity at \SI{3.6}{\micro\metre} due to \SI{3.4}{\micro\metre} and \SI{3.9}{\micro\metre} absorptions & $1 - \left( \frac{a \cdot R3450 + b \cdot R3875}{R3610} \right)$ & Carbonates will be $>$ background values $>$ 0  \\ \hline
ISLOPE1 & spectral slope 1 & $\frac{R1815 - R2530}{W2530 - W1815}$ & Ferric coating on dark rocks  \\ \hline
ICER1\_2 & \ce{CO2} and \ce{H2O} ice band depth & $1 - \left( \frac{1 - BD1435}{1 - BD1500_2} \right)$ &  \ce{CO2} and \ce{H2O} ice mixtures; $>$ 1 for more \ce{CO2} and $>$ 1 for more \ce{H2O} \\ \hline
D2200 & \SI{2.2}{\micro\metre} dropoff & $1 - \left( \frac{\frac{R2210}{RC2210} + \frac{R2230}{RC2230}}{2 \cdot \frac{R2165}{RC2165}} \right)$\footnote{Slope for RC\#\#\#\# anchored at R1815 and R2430} & \ce{Al-OH} minerals  \\ \hline
D2300 & \SI{2.3}{\micro\metre} dropoff & $1 - \left( \frac{\frac{R2290}{RC2290} + \frac{R2320}{RC2320} + \frac{R2330}{RC2330}}{\frac{R2120}{RC2120} + \frac{R2170}{RC2170} + \frac{R2210}{RC2210}} \right)$\footnote{Slope for RC\#\#\#\# anchored at R1815 and R2530} & Hydroxylated \ce{Fe}, \ce{Mg} silicates strongly $>$ 0  \\ \hline
BD1900r2 & \SI{1.9}{\micro\metre} \ce{H2O} band depth & $1 - \left( \frac{\frac{R1908}{RC1908} + \frac{R1914}{RC1914} + \frac{R1921}{RC1921} + \frac{R1928}{RC1928} + \frac{R1934}{RC1934} + \frac{R1941}{RC1941}}{\frac{R1862}{RC1862} + \frac{R1869}{RC1869} + \frac{R1875}{RC1875} + \frac{R2112}{RC2112} + \frac{R2120}{RC2120} + \frac{R2126}{RC2126}} \right)$ & \ce{H2O}  \\ \hline
LCPINDEX2 & Detect broad absorption centered at \SI{1.8}{\micro\metre}  & $RB1690 \cdot 0.20 + RB1750 \cdot 0.20 + RB1810 \cdot 0.30 + RB1870 \cdot 0.20$\footnote{Slope for RC\#\#\#\# anchored at R1560 and R2450} & Pyroxene  \\ \hline
HCPINDEX2 & Detect broad absorption centered at \SI{2.12}{\micro\metre} & $RB2120 \cdot 0.10 + RB2140 \cdot 0.10 + RB2230 \cdot 0.15 + RB2250 \cdot 0.30 + RB2430 \cdot 0.20 + RB2460 \cdot 0.15$\footnote{Slope for RC\#\#\#\# anchored at R1690 and R2530}  & Pyroxene  \\ \hline
OLINDEX3 & Detect broad absorption centered at \SI{1}{\micro\metre} & $RB1080 \cdot 0.03 + RB1152 \cdot 0.03 + RB1210 \cdot 0.03 + RB1250 \cdot 0.03 + RB1263 \cdot 0.07 + RB1276 \cdot 0.07 + RB1330 \cdot 0.12 + RB1368 \cdot 0.12 + RB1395 \cdot 0.14 + RB1427 \cdot 0.18 + RB1470 \cdot 0.18$ \footnote{Slope for RC\#\#\#\# anchored at R1750 and R2400} & Olivines  \\ \hline
%HCPINDEX3\footnote{The HCPINDEX3 formulation has been obtained via personal communication of J. Flahaut.} & Detect broad absorption centered at \SI{2.12}{\micro\metre} & TBD & Pyroxene \\ \hline
\caption{List of currently available CRISM products in PlanetServer. The table has been adapted from \cite{Viviano-Beck2014}. Ongoing work will include the totality of the CRISM products in \cite{Viviano-Beck2014}.}
\label{tab:products}
\end{longtable}
\end{landscape}

The $a$ and $b$ parameters in Table \ref{tab:products} are weighted parameters used to accurately represent the band depth of asymmetric absorption features when $\lambda_C$ (center wavelength) is not centered
between $\lambda_S$ (short wavelength) and $\lambda_L$ (long wavelength). They are calculated as follows:
\begin{gather}
b = \frac{\lambda_C - \lambda_S}{\lambda_L - \lambda_S}\\
a = 1 - b
\end{gather}

\subsubsection{Spectral Analysis tool}
\label{sec:spectral}

Spectra can be easily analyzed in PlanetServer. The Spectral Analysis tool is automatically triggered when clicking in a location of an image (in the web client the user needs to firstly open the designated dock). The tool collects the latitude and longitude of the clicked point, transform it into the same units as in the Coordinate Reference System (CRS) of the image and build a WCPS query to retrieve the spectral data. The server will produce a comma-separated file (CSV) that will be read by the tool. If the tool is used in the web client,  the spectra will be shown in the plot dock and the user will be able to zoom in and out and load different laboratory spectra available in the PDS Geosciences Spectral Library\footnote{Available in: http://speclib.rsl.wustl.edu} or the USGS splib06a (\cite{Clark2007}) spectral library. The user can load one single spectra or collect multiple spectra. The multiselection of spectra is shown in different colors, so each spectra color corresponds to one location. The spectra collection is not reduced to one image allowing the user to collect data among different images. In case the tool is used in the Python API, a new window will be created each time a location is clicked. The API can currently analyze spectra in a single image at a time and spectral libraries are not yet automatically implemented although it is possible to add them manually within a few lines of code.

The tool is also used for spectral ratio calculations (currently available only in the web client). The spectral ratio is located in the ratio dock. Once the ratio dock is opened the user can select the denominator and numerator and click in the image to collect the spectra. A ratio is automatically computed and displayed. Both plots can be downloaded in different formats for further analysis outside PlanetServer.

\subsection{Datasets}

As Table \ref{tab:datasets} illustrates, PlanetServer contains three different types of data: Base maps, DTMs and hyperspectral images of Mars and the Moon. The Mars client includes a global Viking and colored Mars Orbiter Laser Altimeter (MOLA) shaded relief used as base map. The Moon client contains a global Lunar Reconnaissance Orbiter (LRO) WAC and a colored LOLA shaded relief base map. Base maps are served using the Web Mapping Service (WMS) OGC standard in both clients. While Viking and LRO WAC base maps endpoints are located in our servers, MOLA and LOLA colored shaded relief base maps endpoints are located in the United States Geological Survey (USGS) servers. The main reason of having different endpoints is to test and prove interoperability between different endpoints of PlanetServer. 

Mars' DTM is a global MOLA DTM served via WMS from GeoServer. The DTM deployed in the Moon client is a global LOLA DTM. Both DTMs are served from GeoServer as WorldWind only accepts BIL format to deploy the DTM and, at the moment, Rasdaman do not support such format. The original DTMs were transformed to BIL format using GDAL.

CRISM is an spectrometer on board NASA's Mars Reconnaissance Orbiter (MRO). CRISM Targeted Reduced Data Record (TRDR) covers visible to infrared spectra, from 0.362 to 3.92 $\mu m$. Each scene is split in two different images, one covering the range from 0.362 to 1 $\mu m$ (S observations) and the other from 1.035 to 3.92 $\mu m$ (L observations). CRISM TRDR images have been acquired with a detector temperature of $\sim$124$^{\circ}$K. M3 is the image spectrometer on board Chandrayaan-1 Lunar Orbiter (CH-1), an Indian Space Research Organization Lunar orbiter. It covers the spectra from 0.43 to 3 $\mu m$. M3 images have been acquired with a detector temperature of $\sim$167$^{\circ}$K Both datasets are downloaded from the PDS archive and processed in our servers before ingesting them in Rasdaman (the process is discussed in section\ref{sec:process-data}).

In Mars, we also serve the CRISM Multispectral Reduced Data Records (MRDR) dataset (currently in beta version). This dataset is a global tiled coverage of Mars build by mosaicking CRISM TRDR images. The dataset has been spectrally downsampled and separated in 3 different products: Spectral data images, false color images, and CRISM products images. In PlanetServer we currently serve only the CRISM products images. These cubes contain 43 bands with a computed CRISM product in each band. We use WCPS in order to select the bands in the RGB combination tool. Currently, we are testing the dataset behavior in PlanetServer. 

In order to do spectral analysis in PlanetServer, the splib06a \cite{Clark2007} laboratory spectra library has been added to the service. The library contains a set of spectral profiles that can be loaded in the Spectral Analysis tool and compared to CRISM spectral profiles. In both clients, we added vectorial data showing the Moon's and Mars' Gazetteer and a list of landing sites.

\begin{sidewaystable}
\begin{table}[H]
\centering
\caption{List of currently available datasets on PlanetServer.}
\label{tab:datasets}
\begin{tabular}{|lllllllll|}
\hline
\rowcolor[HTML]{C0C0C0}
Body & Dataset & Source & Level & Nº of cubes & Dataset size & Cube size & Pixel size & Retrieval \\ \hline
\rowcolor[HTML]{EFEFEF}
\hline
Mars &  &  &  &  &  &  &  & \\ \hline
 & Base map & Viking & L4 & 1 & 12 GB & 12 GB & 233m/pix & WMS\\ \hline
 & Base map\footnote{As this base map is hosted at the USGS server we do not include the size of the dataset} & MOLA & L4 & 1 & -- & -- & 463m/pix & WMS\\ \hline
 & DTM & Mola global DTM & L4 & 1 & 2GB & 2GB & 463m/pix & WMS\\ \hline
 & Hyperspectral Cubes & CRISM TRDR & L3 & 21659 & 9.6 TB & 50 - 250 MB & 20m/pix & WCPS\\ \hline
 & CRISM products & CRISM MRDR & L4 & 99 & 21 GB & 170 - 300 MB & 232m/pix &WCPS\\ \hline
 & Spectral laboratory data & splib06a & L4 & 1 & -- & -- & -- &WCPS\\ \hline
 & Gazetteer & -- & L4 & 1 & -- & -- & -- &WorldWind\\ \hline
 & Landing sites & -- & L4 & 1 & -- & -- & -- &WorldWind \\ \hline
\rowcolor[HTML]{EFEFEF}
 \hline
Moon &  &  &  &  &  &  &  &\\ \hline
 & Base map & LRO WAC & L4 & 1 & 1.4GB & 1.4GB & 100m/pix &WMS\\ \hline
 & Base map\footnote{As this base map is hosted at the USGS server we do not include the size of the dataset} & LRO WAC & L4 & 1 & -- & -- & 118m/pix &WMS\\ \hline
 & DTM & LOLA & L4 & 1 & 8GB & 8GB & 118m/pix & WMS\\ \hline
 & Hyperspectral cubes & M3 & L2b & 333 & 3.2 TB & 500MB - 3GB & 110m/pix & WCPS\\ \hline
 & Gazetteer & -- & L4 & -- & -- & -- & -- & WorldWind\\ \hline
 & Landing sites & -- & L4 & -- & -- & -- & -- & WorldWind\\ \hline
\end{tabular}
\end{table}
\end{sidewaystable}

\subsubsection{Data processing}
\label{sec:process-data}
CRISM and M3 data have their particular processing pipeline before they can be ingested into rasdaman and be used in PlanetServer, and some rasdaman features need to be created in order to use non-Earth planetary science data. Although both pre-process pipelines have some similarities, each have special steps that need to be addressed. As stated in the introductory section both datasets are publicly available through the PDS archive, therefore obtaining the data is as simple as fetching the data from such archives.

In order to configure Rasdaman to work with the mentioned datasets, we created the CRS for the Moon and Mars and declared them in the Semantic Coordinate Reference System Resolver (SECORE)(\cite{Misev2012}), a server which resolves CRS URLs into full CRS definitions represented in Geographic Markup Language (GML) 3.2.1. The CRS and SECORE will be further discussed in section \ref{sec:crs}

\paragraph{CRISM}
\label{sec:crism}
We created a set of scripts to download  CRISM V3 L2 data containing reflectance information, which search the entire CRISM catalog in the PDS archive. The script creates a file containing links to the available images. The file is later executed using wget in order to download the images. The search script can be used the first time to build the database and subsequently to update the existing database. In the later case it will check the existing list of images in PlanetServer and compare to the list in the PDS archive. In case new data are available, the script will create a new download list. The script can be set to run automatically as a cron job or be launched manually at any time. Currently, PlanetServer's download scripts are launched manually twice a year or after a major update of the PDS archive has been announced. The downloading scripts are available in our GitHub repository. The data are stored according to the volume they belong to (e.g mrocr\_2010, mrocr\_2011, etc). The data are then processed using CAT routines on ENVI. Several CAT IDL routines have been altered to automate the process and make them non-interactive, allowing us to run the process in the background. The result are atmospherically corrected and map projected CRISM images. The output IMG files are transformed into GeoTIFF using GDAL in order to be ingested into rasdaman. In the transformation process the metadata is kept in its original format in order to be used later by the ingestion script. Ingesting data into rasdaman can be easily performed by creating a so called ingredient(a json structured file descriptor) (Listing \ref{lst:ingredient}) and using the built-in routine wcst\_import. In the ingredient, information related to the CRS, WMS capabilities, band nomenclature and null values are included. The wcst\_import routine will read the ingredient parameters and depending on those ingest the data in one of the several possible ways. In this case a map mosaic mode ingredient with WMS enabled has been chosen. We also renamed the bands by a more convenient nomenclature as the name will be used when building the WCPS queries . Once the ingredients are created, the data are ingested into petascope. This data can be easily queried using either WMS, WCS or WCPS OGC standards.

\paragraph{M3}
\label{sec:m3}
M3 data are also downloaded directly from PDS using a set of scripts. In this case we downloaded L1B and L2 data. L2 data contains thermally corrected reflectance data which is the data we are interested in. The reason of downloading both data levels is that the L2 data needs to be re-projected using Integrated Software for Imagers and Spectrometers (ISIS) (\cite{Gaddis1997} and \cite{Torson1997}) routines and they are currently only available for L1B data. It appears that the reflectance file has the same characteristics (lines, samples, bands, bit type) as the M3 L1B data. A method has been found to change the L1B data label file to point into the L2 reflectance data (Hare, T., personal communication (2016)). This allows us to run the ISIS routine "cam2map" to reproject L2 data by using the L1B data label file. The pipeline script containing all the necessary ISIS routines is launched in order to obtain the L2 map projected data. The data is transformed to GeoTIFF using GDAL for the same reason as in section \ref{sec:crism}. The only differences in the M3 ingestion ingredients are the CRS URL (pointing to Moon based CRSs) and the number of bands. As soon as higher level and/or re-calibrated data are available, they will be added or substitute the current data products.

\begin{figure}[H]
\centering
\includegraphics[width=300pt]{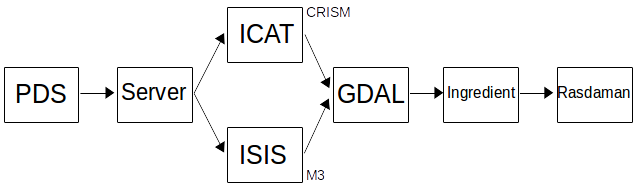}%ext=bmp,jpg,...
\caption[Pipeline for processing CRISM and M3 data.]{Pipeline for processing CRISM and M3 data. Both pipelines share some steps such as the GeoTIFF transformation while some parts of the pipeline are dataset tied.}
\label{fig:pipeline}
\end{figure}

\paragraph{Planetary CRS}
\label{sec:crs}

Allowing full support of non-terrestrial CRS in web-based GIS is an ongoing effort among planetary scientists. Some proposals have been carried out (\cite{Rossi2016}) although no CRS have been yet included. Currently, the official CRS is defined by the International Astronomical Union (IAU) in 2000, named MARS2000. The MARS2000 definition consists of an ellipsoid with a 3396190 meter semi-major axis and a 3376200 meter semi-minor axis. As this definition is planetocentric, different from the terrestrial CRS which are planetographic, MARS2000 is not to be supported by many GIS clients (\cite{Oosthoek2014}). In order to solve this issue we defined the Martian CRS as a 3396190 meters ellipsoid with no flattening.

SECORE is the CRS resolver used by rasdaman. The CRS is developed in GML 3.2.1, validated and included into SECORE using its front-end web version, which will create a unique URL pointing to the CRS. The URL is included in the ingredients in order to define the CRS of the images while ingesting into rasdaman. For earth-based data, most of the CRS used are stored in opengis.net with their own unique URL. We created the CRS needed in PlanetServer and included them in SECORE. We defined ten CRS: 4 for Mars and 6 for the Moon. The list of CRSs and their unique URL can be found in  Table \ref{tab:crs} in the Appendix.

\subsection{Examples}
\label{sec:examples}

Mineral characterization in Mars is an ongoing research. Several studies of deposition and alteration phases have been carried out in different locations of the planet (\cite{Mustard2008}, \cite{Carter2013}, \cite{Bibring2005}). We carried out a study of the characterization of chlorite, prehnite and kaolinite in the Nili Fossae area, thus allowing us to compare the results obtained with PlanetServer to previous studies (\cite{Ehlmann2009} and \cite{Viviano-Beck2014}).

\subsubsection{Chlorite and Prehnite}
\label{sec:chlo}
Chlorites are phyllosilicates formed by hydrothermal, metamorphic and diagenetic reactions (\cite{Ehlmann2009}). Prehnite, which usually appears in association with chlorite and pumpellyite, is a calcium aluminum silicate hydroxide material formed after hydrothermal or metamorphic activity. Chlorite shows a strong absorption band centered at 2.33 - \SI{2.35}{\micro\metre}  which combined with a shoulder at \SI{2.26}{\micro\metre} distinguishes chlorites from serpentines. Prehnite shows also an absorption band centered at 2.35 - \SI{2.36}{\micro\metre} coinciding with Fe-rich chlorites. An absorption band at \SI{1.48}{\micro\metre}  is uniquely related with the presence of prehnite. 

\begin{figure}[H]
\centering
\includegraphics[width=\textwidth]{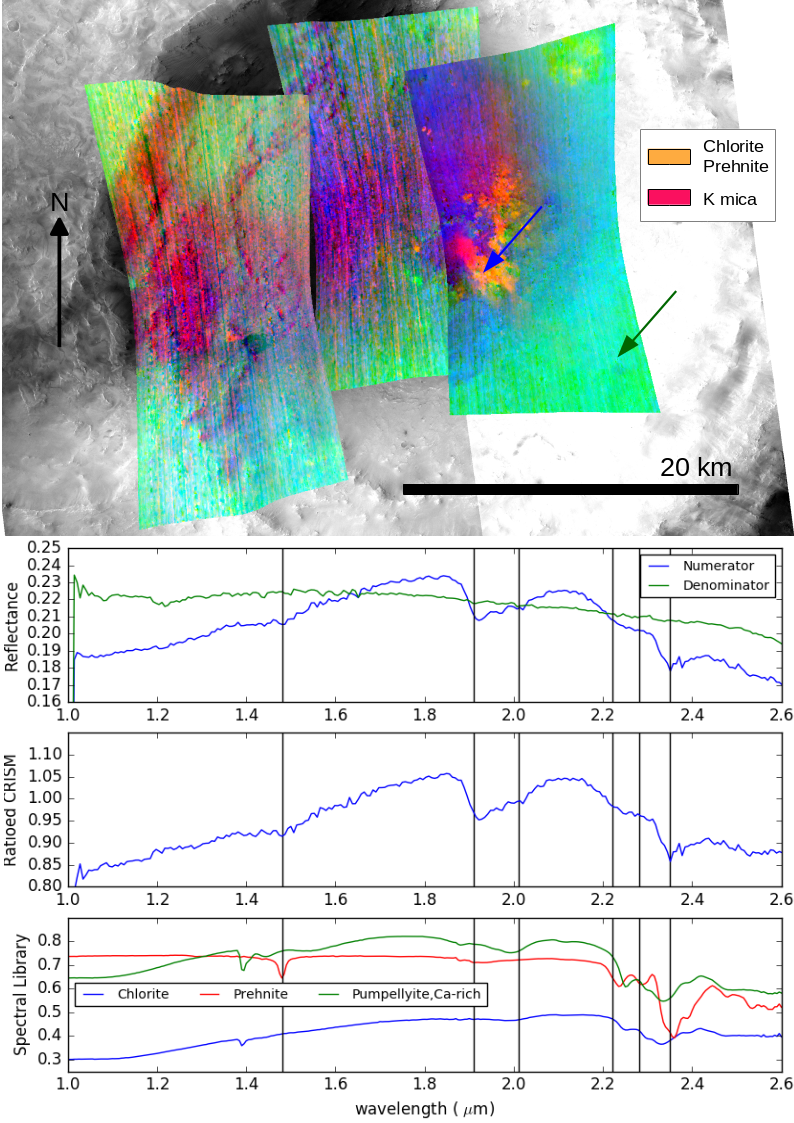}%ext=bmp,jpg,...
\caption[Chlorite around Nili Fossae.]{Chlorite in Toro crater within Nili Fossae. The RGB combination has been calculated using R: BD2355; G: D2300; B: MIN2200 which highlights the possible presence of chlorite in yellow. Raw spectra (Figure \ref{fig:chlorite}A) has been collected from the image FRT0000B1B5. The numerator (blue arrow) is located at 17.00813$^{\circ}$N, 71.85837$^{\circ}$E and the denominator (green arrow) at 16.91249$^{\circ}$N, 71.99203$^{\circ}$E. The ratioed spectra is shown in Figure \ref{fig:chlorite}B. Spectral analysis (Figure \ref{fig:chlorite}C) shows a mix of chlorite and prehnite with absorption bands at \SI{1.48}{\micro\metre}, \SI{2.22}{\micro\metre}, \SI{2.28}{\micro\metre} and \SI{2.35}{\micro\metre}. A clear mix of prehnite with hydrated minerals is present hence the absorption band at \SI{1.9}{\micro\metre}. The shape of the continuum between \SI{2.2}{\micro\metre} and \SI{2.3}{\micro\metre} and after \SI{2.5}{\micro\metre} suggests a putative mixture with either chlorite and/or pumpelliyte.}
\label{fig:chlorite}
\end{figure}

Figure \ref{fig:chlorite} shows a crater in the Nili Fossae area (17$^{\circ}$N, 72$^{\circ}$E) with a high chlorite content region. In \cite{Ehlmann2009} the combination used to characterize chlorite and prehnite (red: BD2350; green: D2300;blue: BD2200) have been derived from \cite{Pelkey2007} and show a distribution of chlorite/prehnite and K mica materials. In our study the images are obtained as a combination of three products (red: BD2355; green: D2300; blue: MIN2200) from \cite{Viviano-Beck2014} which are equivalent products. Chlorite and prehnite materials are shown in  yellow in Figure \ref{fig:chlorite} and are mainly located  in the central part of the crater.

The spectral analysis, shown in Figure \ref{fig:chlorite}A, B and C, corroborates the presence of chlorites and prehnites. Figure \ref{fig:chlorite}A shows in blue the numerator (using a 3x3 kernel) where the absorption bands at \SI{1.9}{\micro\metre} and \SI{2.35}{\micro\metre} are already very prominent. The ratioed CRISM graph (Figure \ref{fig:chlorite}B) enhances the above mentioned absorption bands. Absorption bands at 1.48, 1.9 and \SI{2.35}{\micro\metre} are associated with the presence of chlorite and prehnite, therefore we can say that the analyzed area contains chlorite, prehnite and pumpellyite. The graphs are compared against laboratory samples spectra included in the spectral library splib06a (\cite{Clark2007}) in order to validate the spectral analysis (Figure \ref{fig:chlorite}C).

\subsubsection{Kaolinite}

Kaolinites are a clay mineral including kaolinite, dickite, nacrite, montmorillonite and halloysite. Kaolinite is formed as a result of hydrothermal alteration and pedogenesis while halloysite is formed as a result of hydrothermal or weathering processes. Kaolinites show a strong absorption band at \SI{1.4}{\micro\metre} and  \SI{2.2}{\micro\metre}. An absorption band at \SI{1.9}{\micro\metre} is also present but shows a variation in strength. Nacrite and Dickite show a distinct absorption band at \SI{2.17}{\micro\metre}. An absorption band at \SI{1.92}{\micro\metre} is also visible which corresponds to halloysites.

\begin{figure}[H]
\centering
\includegraphics[width=\textwidth]{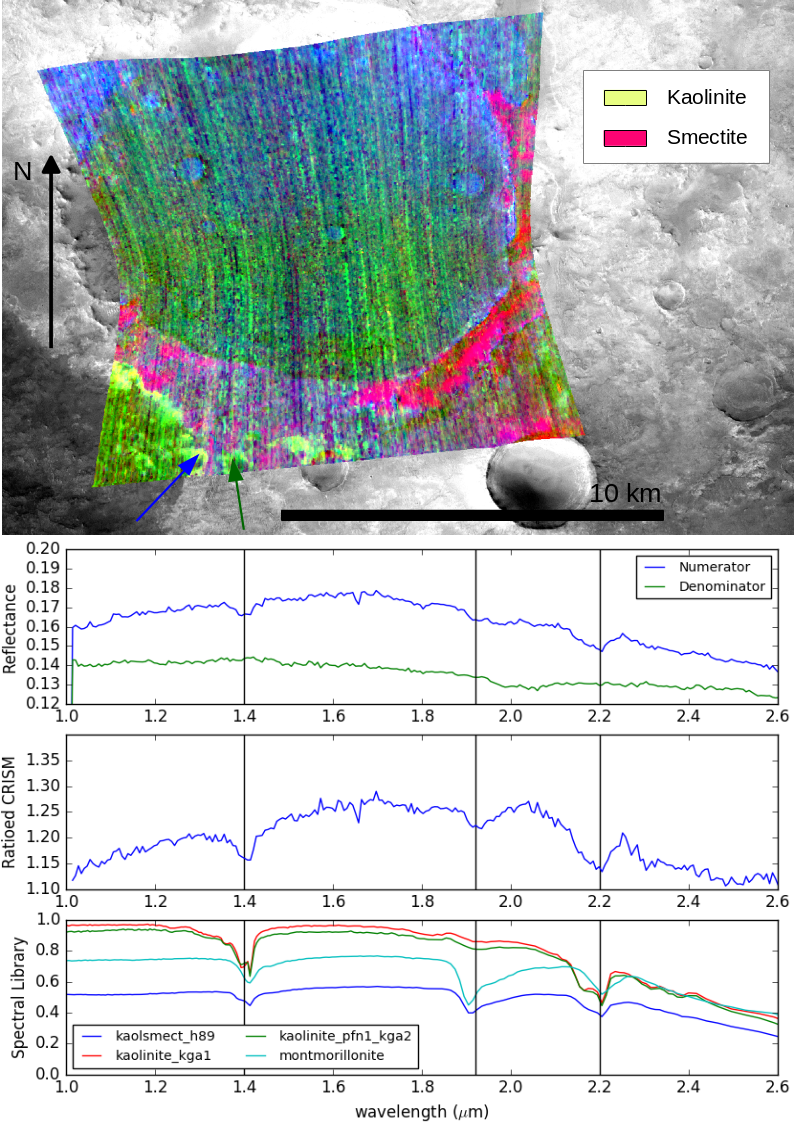}%ext=bmp,jpg,...
\caption[Kaolinite around Nili Fossae.]{Kaolinite found in the image FRT0000C202 located in Nili Fossae. The RGB combination has been calculated using R: BD1900\_2; G: MIN2200; B:D2300. A) Spectra collected at 21.17468$^{\circ}$N, 73.31922$^{\circ}$E (numerator, blue arrow) and 21.16931$^{\circ}$N, 73.33749$^{\circ}$E (denominator, green arrow) is shown. The ratioed (B) spectra of the numerator and denominator in A is presented. The vertical lines at 1.4, 1.92, and 2.2 mm correspond to absorption bands related with kaolinite, halloysite or kaolinite smectite clays. As seen in C the CRISM spectra has a better fit  to kaolinite rather than montmorillonite as it coincides with the characteristic \SI{2.16}{\micro\metre} absorption band found in kaolinites.}
\label{fig:kaol}
\end{figure}

Kaolinite is not as widely detected as chlorites but some traces of it can be found in Nili Fossae (\cite{Brown2010a} and \cite{Ehlmann2009}) and other areas (\cite{Mustard2008b} and \cite{Cuadros2013}). Figure \ref{fig:kaol} shows a small crater of around 5km diameter with a high content of Kaolinite (yellow) in its south-easternmost part of the rim (21$^{\circ}$N, 73$^{\circ}$E). The images are obtained after adapting the CRISM products from \cite{Pelkey2007} used in \cite{Ehlmann2009} (red: BD1900H; green: BD2200; blue: D2300) to the more actual developed by Viviano-beck (red: BD1900\_ 2; green: MIN2200; blue:D2300). Such products are used to characterize kaolinite and smectite materials.

Smectite is shown in pink while kaolinites are shown in yellow and light-green. Note the high noise content present in Figure \ref{fig:kaol} which is also shown in a more dark green. It is mainly due to the fact that the absorption band corresponding to the green channel is near to the noise level in CRISM images. Applying image processing techniques such as calculating the median and dividing by it to reduce the noise will give us a less noise image but will also reduce the absorption band at \SI{2.17}{\micro\metre}. We decided to use the CRISM products without post-processing the image and further investigate the consequences of applying such filters in the results.

Figure \ref{fig:kaol}A shows in blue the  reflectance of the numerator where already absorption bands at 1.4 and \SI{2.2}{\micro\metre} can be easily recognized. The ratioed CRISM (Figure \ref{fig:kaol}B) increases its signal making it easier to recognize the absorption band at \SI{1.92}{\micro\metre}. The above mentioned absorption bands are tied to the presence of kaolinite, smectite, montmorillonite, nacrite, dickite and halloysite. As in the section \ref{sec:chlo},  all the results are compared against laboratory samples included in the spectral library splib06a which can be seen in Figure \ref{fig:kaol}C.

\section{Discussion}

PlanetServer is presented in this work as a work in process open-source and reproducible alternative to visualize and analyze hyperspectral images. The main use case is linked to Mars, but the applicability of code, tools and workflow is general purpose enough to be applied to other Solar System bodies and Earth Observation. The cross validation and the scientific results show the robustness of PlanetServer; further validation is planned. The scientific results are compared to previous studies (\cite{Ehlmann2009}, \cite{Viviano-Beck2014}. Same minerals have been characterized with positive results. All the process is pursued in the same environment without having to know specifically any of the OGC standards, WCPS specially. As the client (\cite{MarcoFiguera2016a}) is open-source and available in GitHub\footnote{https://github.com/planetserver/ps2-www-client}, it is easy to modify and add more CRISM products and WCPS queries. 

As all the products can be downloaded, the results obtained in PlanetServer (images and plots) can be easily included in scientific publications. Giving the fact that PlanetServer and rasdaman Community edition\footnote{http://rasdaman.org/} are open-source, both the client and the database manager can be deployed in another system in order to serve different datasets. PlanetServer contributes towards cost-effective workflows as it is not required to pay any licenses for a fully functional setup. 

\subsection{Ongoing work}

Ongoing work includes the development of a Python API. The API will help in the creation of all the CRISM products mentioned in \citep{Viviano-Beck2014}. It will only need four input arguments: Coverage Identification and the name of 3 CRISM products to be stored in each channel. The output will be an image with the given parameters. The inclusion of spectral analysis in the API is being evaluated. Once the API is published we will provide a set of Jupyter Notebooks in order to give an overview of its capabilities. An early stage of PlanetServer working on a Jupyter notebook is already accessible\footnote{https://github.com/earthserver-eu/INSPIRE-notebooks/blob/master/ps2\_inspire\_notebook.ipynb}.

New datasets are being considered. The CRISM MRDR dataset (\cite{McGuire2013}) will be shortly included in PlanetServer and will allow the possibility to analyze bigger areas with one WCPS query. The CRISM MRDR dataset is composed by a set of mosaics served as tiles covering the whole Martian surface\footnote{http://pds-geosciences.wustl.edu/missions/mro/090220\_current\_if\_tru.png}. 

New M3 data is being processed to be ingested in the Lunar client. The reason of using the pipeline described in section \ref{sec:m3} is to apply the thermal correction of L2 data into L1B data and process the data using ISIS routines. We found out that, while some images are processed without any error, some images appear to be corrupted during the process. Further investigations are being taken in order to analyze and solve the issue. An hypothesis of the error source could be the hacking of the label files which corrupts the ISIS3 procedure. Once the M3 data is processed we will achieve a nearly global M3 coverage. 

The inclusion of topographic datasets is also being evaluated and tested using High Resolution Stereo Camera (HRSC, Mars Express) data (\cite{Jaumann2015}). PlanetServer will include a set of DTMs for all the planetary bodies as well as a set of tools to pursue topographic analysis - traverse, roughness, slopes and aspects. A possible application of including topographic analysis to the current tools is landing site characterization (\cite{Golombek2007} and \cite{Golombek2003}). The Java Mission-planning and Analysis for Remote Sensing (JMARS) (\cite{Christensen2009}) is a software developed in Java allowing the user to plan and process landing site selection analysis. While JMARS is a very powerful tool it needs to be downloaded and run locally. Including the topographic tools, a set of illumination maps (\cite{MarcoFiguera2014} and \cite{MarcoFiguera2016}) and the existing analytical tools will grant PlanetServer to be an online alternative for Landing site selection. 

Discussions have started in order to evaluate the inclusion of OMEGA data for Mars and Messenger data for Mercury. Future plans for the inclusion of a hyperspectral clustering tool in PlanetServer using a server based python API is being evaluated. The inclusion of this feature needs to be further evaluated as it will increase the stress in the servers.

Efforts are being made in order to standardize the declaration of IAU CRSs and include them in opengis (\cite{Rossi2016}).

\section{Conclusions}

The results obtained in this research demonstrates PlanetServer as a reliable tool for the visualization and analysis of hyperspectral data. Several investigations have been pursued in order to validate our products giving a positive result in all the tests. It has been proved the reliability of PlanetServer in the analysis of spectra and the visualization of images and band math combinations. The fact of having an all-in-one platform provides us the possibility of running quick tests in a very short time. As an example, the cross validation of the chlorites and prehnites products was completed in just one hour based on the investigation pursued by \cite{Ehlmann2009}. Once the cross validation of the chlorites was finished, finding new areas with chlorite and prehnite was a routine process comparable to those done by using other software such as ENVI.

PlanetServer can be easily integrated in Python providing us the possibility to make use of all the libraries present in Python. This extends the power of PlanetServer for the analysis of CRISM images. The python API can be easily integrated and combined with existing geo-libraries such as gdal, providing the possibility of map mosaic the results.

The combination of OGC standards, open-source server and client tools as well as openly available algorithms together with WCPS versions of hyperspectral formulas allows reproducibility of scientific observations and surface mapping. We need a joint effort including universities, companies and space agencies in order to lead the development of new tools into a more openly and easily accessible philosophy. It is proved that by making accessible tools to analyze planetary data we can reduce time and budget spent in developing custom tools and focus on the analysis and the scientific output.

\section*{Acknowledgments}
The EarthServer-2 project receives funding from the European Union’s Horizon 2020 research and innovation program under the grant agreement No 654367 (e-Infrastructures).
The authors wish to thank the CLAMV team (Achim Gelessus, Florian Neu) to let us  and help us into using the cluster, the L-sis group (Vlad Merticariu, Alex Dumitru and Dimitar Misev) for all their help in rasdaman related issues. We thank Melissa Martinot for helping finding laboratory spectra samples. We would like to thank Trent Hare for its help creating the M3 pipeline and for a all the information regarding CRSs. Also, we would like to thank the EarhtServer-2 Consortium as a whole and particularly its Service Partners for the valuable discussions and for their help.

\section{Appendix}
\label{sec:appendix}

Table \ref{tab:crs} shows a list of all the available CRS created in PlanetServer and stored in SECORE. The CRS are created in XML format and added to SECORE using its web interface. The CRS's correctness is checked by SECORE prior ingestion.

\begin{sidewaystable}
\centering
\caption{List of URLs of available CRSs in PlanetServer}
\label{tab:crs}
\begin{tabular}{|l|l|l|}
\hline
\rowcolor[HTML]{C0C0C0}
Body & CRS & URL \\ \hline
\rowcolor[HTML]{EFEFEF}
\hline
Mars &  &  \\ \hline
 & Mars Equirectangular & \texttt{http://access.planetserver.eu:8081/def/crs/PS/0/Mars-equirectangular} \\ \hline
 & Mars Geographic &  \texttt{http://access.planetserver.eu:8081/def/crs/PS/0/Mars-geographic}\\ \hline
 & Mars North polar stereographic & \texttt{http://access.planetserver.eu:8081/def/crs/PS/0/Mars-stereographic-north} \\ \hline
 & Mars South polar stereographic & \texttt{http://access.planetserver.eu:8081/def/crs/PS/0/Mars-stereographic-south
} \\ \hline
\rowcolor[HTML]{EFEFEF}
 \hline
Moon &  &  \\ \hline
 & Moon Equirectangular & \texttt{http://access.planetserver.eu:8081/def/crs/PS/0/Moon-equirectangular
} \\ \hline
 & Moon Geographic & \texttt{http://access.planetserver.eu:8081/def/crs/PS/0/Moon-geographic
} \\ \hline
 & Moon North polar stereographic & \texttt{http://access.planetserver.eu:8081/def/crs/PS/0/Moon-gnomonic-north
} \\ \hline
 & Moon South polar stereographic & \texttt{http://access.planetserver.eu:8081/def/crs/PS/0/Moon-gnomonic-south
} \\ \hline
 & Moon North polar gnomonic & \texttt{http://access.planetserver.eu:8081/def/crs/PS/0/Moon-stereographic-north
} \\ \hline
 & Moon South polar gnomonic & \texttt{http://access.planetserver.eu:8081/def/crs/PS/0/Moon-stereographic-south
} \\ \hline
\end{tabular}
\end{sidewaystable}

Table \ref{tab:locationsl} shows the location in degrees of all the locations used in section \ref{sec:examples}. Images below 65º use equirectangular projection and above 65º polar stereographic projection. 
% \usepackage{multirow}
\begin{table}[H]
\centering
\caption{Location of Numerators and Denominators used in Figures 7 and 8}
\label{tab:locationsl}
\begin{tabular}{|l|l|l|l|}
\hline
\rowcolor[HTML]{C0C0C0}
\multicolumn{1}{|c|}{\cellcolor[HTML]{C0C0C0}{\color[HTML]{333333} }} & \multicolumn{2}{c|}{\cellcolor[HTML]{C0C0C0}{\color[HTML]{333333} Location}} & \multicolumn{1}{c|}{\cellcolor[HTML]{C0C0C0}{\color[HTML]{333333} }} \\ \cline{2-3}
\rowcolor[HTML]{C0C0C0}
\multicolumn{1}{|c|}{\multirow{-2}{*}{\cellcolor[HTML]{C0C0C0}{\color[HTML]{333333} Label}}} & \multicolumn{1}{c|}{\cellcolor[HTML]{C0C0C0}{\color[HTML]{333333} Latitude}} & \multicolumn{1}{c|}{\cellcolor[HTML]{C0C0C0}{\color[HTML]{333333} Longitude}} & \multicolumn{1}{c|}{\multirow{-2}{*}{\cellcolor[HTML]{C0C0C0}{\color[HTML]{333333} Image ID in PlanetServer}}} \\ \hline
\rowcolor[HTML]{EFEFEF}
Figure 6 &  &  &  \\ \hline
Numerator & 17.00813 & 71.85837 & frt0000b1b5\_07\_if165l\_trr3 \\ \hline
Denominator & 16.91249 & 71.99203 & hrl000086ca\_07\_if182l\_trr3 \\ \hline
\rowcolor[HTML]{EFEFEF}
Figure 7 &  &  &  \\ \hline
Numerator & 21.17468 & 73.31922 & frt0000c202\_07\_if165l\_trr3 \\ \hline
Denominator & 21.16931 & 73.33749 & frt0000c202\_07\_if165l\_trr3 \\ \hline
\end{tabular}
\end{table}

Table \ref{tab:lab} shows a detailed list of the laboratory spectra used to analyze the CRISM spectra in section \ref{sec:examples}.

\begin{table}[H]
\centering
\caption{List of sample names and sources used in Figures 4-7. }
\label{tab:lab}
\begin{tabular}{|l|l|l|}
\hline
\rowcolor[HTML]{C0C0C0}
\multicolumn{1}{|c|}{\cellcolor[HTML]{C0C0C0}Label} & \multicolumn{1}{c|}{\cellcolor[HTML]{C0C0C0}Sample ID} & \multicolumn{1}{c|}{\cellcolor[HTML]{C0C0C0}Source} \\ \hline
\rowcolor[HTML]{EFEFEF}
Figure 7 &  &  \\ \hline
Chlorite & c2cl14\_bdvnir & PDS Geoscience Spectral Library \\ \hline
Prehnite & c1ze03\_bdvnir & PDS Geoscience Spectral Library \\ \hline
Pumpellyite,Ca-rich & c1ze01\_bdvnir & PDS Geoscience Spectral Library \\ \hline
%Pumpellyite,Mg-rich" & c1ze02\_bdvnir & PDS Geoscience Spectral Library \\ \hline
\rowcolor[HTML]{EFEFEF}
Figure 8 &  &  \\ \hline
kaolsmect\_h89 & kaolsmect\_h89fr2.25636 & splib06a \\ \hline
kaolinite\_kga1 & kaolinite\_kga1.12117 & splib06a \\ \hline
kaolinite\_pfn1\_kga2 & kaolinite\_pfn1\_kga2.12176 & splib06a \\ \hline
montmorillonite & montmorillonite\_swy1.14688 & splib06a \\ \hline
%halloysite & halloysite\_nmnh106236.8988 & splib06a \\ \hline
%nacrite & nacrite\_gds88.15578 & splib06a \\ \hline
%dickite & dickite\_nmnh106242.6820 & splib06a \\ \hline
\end{tabular}
\end{table}

%\section{Bibliography styles}
%
%Here are two sample references: \cite{Bibring2004}.
Example of ingredient used for the ingestion of a CRISM image:

\begin{lstlisting}[ language=SQL,
           caption={Ingredient used in PlanetServer to ingest a CRISM image in the Rasdaman database.},
           label=lst:ingredient,
           showspaces=false,
           basicstyle=\footnotesize,
           breaklines=true,
           numbers=left,
           numberstyle=\tiny,
           commentstyle=\color{gray}
        ]
{
  "config": {

    "service_url": "http://localhost:8080/rasdaman/ows",

    "tmp_directory": "/tmp/",

    "crs_resolver": "http://localhost:8080/def/",

    "default_crs": "http://localhost:8080/def/crs/PS/0/Mars-equirectangular",

    "subset_correction": false,

    "mock": false,

    "automated": true
  },
  "input": {

    "coverage_id": "frt00005f1f_07_if163l_trr3",
    "paths": [

     "../frt00005f1f_07_if163l_trr3_CAT_scale_trial_p.img.tif"
    ]
  },
  "recipe": {

    "name": "map_mosaic",
    "options": {

      "tiling": "ALIGNED [0:1023, 0:1023] TILE SIZE 4194304",
        "wms_import": true,
		"band_names": [
"band_1",
"band_2",
"band_3",
"band_4",
"band_5",
.
.
.
"band_438"
	]
    }
  }
}
\end{lstlisting}

A list of different WCPS queries is provided. The following query is used to retrieve spectra profiles from CRISM images:

\begin{lstlisting}[
           language=SQL,
           caption={Single pixel spectra query},
           label=lst:spec,
           showspaces=false,
           basicstyle=\footnotesize,
           breaklines=true,
           %numbers=left,
           %numberstyle=\tiny,
           commentstyle=\color{gray}
        ]
for c in (frt0000a0ac_07_if165l_trr3) return encode(c[ N(1277.941354035707:1277.941354035707), E(1022.353146921535:1022.353146921535) ], "csv")
\end{lstlisting}

The following WCPS query is used to create a false color RGB combination. This query is used as the default RGB combination when an image is selected.

\begin{lstlisting}[
           language=SQL,
           caption={RGB false color WCPS query},
           label=lst:rgbcomb,
           showspaces=false,
           basicstyle=\footnotesize,
           breaklines=true,
           %numbers=left,
           %numberstyle=\tiny,
           commentstyle=\color{gray}
        ]
for data in (frt0000a0ac_07_if165l_trr3) return encode( { 
red: (float)((int)(255 / (max( data.band_233) - min(data.band_233))) * (data.band_233 - min(data.band_233))); 
green: (float)((int)(255 / (max( data.band_13) - min(data.band_13))) * (data.band_13 - min(data.band_13))); 
blue: (float)((int)(255 / (max( data.band_78) - min(data.band_78))) * (data.band_78 - min(data.band_78))) ; 
alpha: (float)((data.band_100 > 0) * 255)}, "png", "nodata=65535")
\end{lstlisting}

The following query creates the HYD browse product from \cite{Viviano-Beck2014}. The first stretching to translate the indexes to the PNG range is added in the query. 
\begin{lstlisting}[
           language=SQL,
           caption={Viviano-beck HYD CRISM product WCPS query},
           label=lst:product,
           showspaces=false,
           basicstyle=\footnotesize,
           breaklines=true,
           %numbers=left,
           %numberstyle=\tiny,
           commentstyle=\color{gray}
        ]
for data in ( frt0000a0ac_07_if165l_trr3 ) return encode( {
Red:(float)((int) ( 255 / ( max((1 - ((1 - (0.607142857))*data.band_171 + (0.607142857)*data.band_213)/(data.band_197))) - min((1 - ((1 - (0.607142857))*data.band_171 + (0.607142857)*data.band_213)/(data.band_197))) )) * ( ((1 - ((1 - (0.607142857))*data.band_171 + (0.607142857)*data.band_213)/(data.band_197))) - min((1 - ((1 - (0.607142857))*data.band_171 + (0.607142857)*data.band_213)/(data.band_197))) ));
Green:(float)((int)( 255 / ( max((1 - ((data.band_173) / ((1 - (0.63125)) * data.band_142 + (0.63125) * data.band_191)))) - min((1 - ((data.band_173) / ((1 - (0.63125)) * data.band_142 + (0.63125) * data.band_191)))) )) * ( ((1 - ((data.band_173) / ((1 - (0.63125)) * data.band_142 + (0.63125) * data.band_191)))) - min((1 - ((data.band_173) / ((1 - (0.63125)) * data.band_142 + (0.63125) * data.band_191)))) ));
Blue:(float)((int)( 255 / ( max((0.5 * (1 - ((data.band_142) / ((1 - (0.36346516)) * data.band_130 + (0.36346516) * data.band_163))) * 0.5 * (1 - ((data.band_151) / ((1 - (0.636167379)) * data.band_130 + (0.636167379) * data.band_163))))) - min((0.5 * (1 - ((data.band_142) / ((1 - (0.36346516)) * data.band_130 + (0.36346516) * data.band_163))) * 0.5 * (1 - ((data.band_151) / ((1 - (0.636167379)) * data.band_130 + (0.636167379) * data.band_163))))) )) * ( ((0.5 * (1 - ((data.band_142) / ((1 - (0.36346516)) * data.band_130 + (0.36346516) * data.band_163))) * 0.5 * (1 - ((data.band_151) / ((1 - (0.636167379)) * data.band_130 + (0.636167379) * data.band_163))))) - min((0.5 * (1 - ((data.band_142) / ((1 - (0.36346516)) * data.band_130 + (0.36346516) * data.band_163))) * 0.5 * (1 - ((data.band_151) / ((1 - (0.636167379)) * data.band_130 + (0.636167379) * data.band_163))))) ));
alpha: (data.band_100 > 0) * 255 }, "tiff", "nodata=65535")
\end{lstlisting}

\bibliography{My_Collection.bib}

\begin{thebibliography}{44}
\expandafter\ifx\csname natexlab\endcsname\relax\def\natexlab#1{#1}\fi
\providecommand{\url}[1]{\texttt{#1}}
\providecommand{\href}[2]{#2}
\providecommand{\path}[1]{#1}
\providecommand{\DOIprefix}{doi:}
\providecommand{\ArXivprefix}{arXiv:}
\providecommand{\URLprefix}{URL: }
\providecommand{\Pubmedprefix}{pmid:}
\providecommand{\doi}[1]{\href{http://dx.doi.org/#1}{\path{#1}}}
\providecommand{\Pubmed}[1]{\href{pmid:#1}{\path{#1}}}
\providecommand{\bibinfo}[2]{#2}
\ifx\xfnm\relax \def\xfnm[#1]{\unskip,\space#1}\fi
%Type = Article
\bibitem[{Aiordǎchioaie and Baumann(2010)}]{Aiordachioaie2010}
\bibinfo{author}{Aiordǎchioaie, A.}, \bibinfo{author}{Baumann, P.},
  \bibinfo{year}{2010}.
\newblock \bibinfo{title}{{PetaScope: An open-source implementation of the OGC
  WCS Geo service standards suite}}.
\newblock \bibinfo{journal}{Lecture Notes in Computer Science (including
  subseries Lecture Notes in Artificial Intelligence and Lecture Notes in
  Bioinformatics)} \bibinfo{volume}{6187 LNCS}, \bibinfo{pages}{160--168}.
\newblock \URLprefix \url{http://dl.acm.org/citation.cfm?id=1876037.1876053},
  \DOIprefix\doi{10.1007/978-3-642-13818-8{\_}13}.
%Type = Article
\bibitem[{Baumann(2010)}]{Baumann2009a}
\bibinfo{author}{Baumann, P.}, \bibinfo{year}{2010}.
\newblock \bibinfo{title}{{The OGC web coverage processing service (WCPS)
  standard}}.
\newblock \bibinfo{journal}{GeoInformatica} \bibinfo{volume}{14},
  \bibinfo{pages}{447--479}.
\newblock \URLprefix \url{http://dl.acm.org/citation.cfm?id=1831146.1831160},
  \DOIprefix\doi{10.1007/s10707-009-0087-2}.
%Type = Article
\bibitem[{Baumann et~al.(1998)Baumann, Dehmel, Furtado, Ritsch and
  Widmann}]{Baumann1998}
\bibinfo{author}{Baumann, P.}, \bibinfo{author}{Dehmel, A.},
  \bibinfo{author}{Furtado, P.}, \bibinfo{author}{Ritsch, R.},
  \bibinfo{author}{Widmann, N.}, \bibinfo{year}{1998}.
\newblock \bibinfo{title}{{The multidimensional database system RasDaMan}}.
\newblock \bibinfo{journal}{ACM SIGMOD Record} \bibinfo{volume}{27},
  \bibinfo{pages}{575--577}.
\newblock \URLprefix \url{http://dl.acm.org/citation.cfm?id=276305.276386},
  \DOIprefix\doi{10.1145/276305.276386}.
%Type = Article
\bibitem[{Baumann et~al.(2015)Baumann, Mazzetti, Ungar, Barbera, Barboni,
  Beccati, Bigagli, Boldrini, Bruno, Calanducci, Campalani, Clements, Dumitru,
  Grant, Herzig, Kakaletris, Laxton, Koltsida, Lipskoch, Mahdiraji, Mantovani,
  Merticariu, Messina, Misev, Natali, Nativi, Oosthoek, Pappalardo, Passmore,
  Rossi, Rundo, Sen, Sorbera, Sullivan, Torrisi, Trovato, Veratelli and
  Wagner}]{Baumann2015}
\bibinfo{author}{Baumann, P.}, \bibinfo{author}{Mazzetti, P.},
  \bibinfo{author}{Ungar, J.}, \bibinfo{author}{Barbera, R.},
  \bibinfo{author}{Barboni, D.}, \bibinfo{author}{Beccati, A.},
  \bibinfo{author}{Bigagli, L.}, \bibinfo{author}{Boldrini, E.},
  \bibinfo{author}{Bruno, R.}, \bibinfo{author}{Calanducci, A.},
  \bibinfo{author}{Campalani, P.}, \bibinfo{author}{Clements, O.},
  \bibinfo{author}{Dumitru, A.}, \bibinfo{author}{Grant, M.},
  \bibinfo{author}{Herzig, P.}, \bibinfo{author}{Kakaletris, G.},
  \bibinfo{author}{Laxton, J.}, \bibinfo{author}{Koltsida, P.},
  \bibinfo{author}{Lipskoch, K.}, \bibinfo{author}{Mahdiraji, A.R.},
  \bibinfo{author}{Mantovani, S.}, \bibinfo{author}{Merticariu, V.},
  \bibinfo{author}{Messina, A.}, \bibinfo{author}{Misev, D.},
  \bibinfo{author}{Natali, S.}, \bibinfo{author}{Nativi, S.},
  \bibinfo{author}{Oosthoek, J.}, \bibinfo{author}{Pappalardo, M.},
  \bibinfo{author}{Passmore, J.}, \bibinfo{author}{Rossi, A.P.},
  \bibinfo{author}{Rundo, F.}, \bibinfo{author}{Sen, M.},
  \bibinfo{author}{Sorbera, V.}, \bibinfo{author}{Sullivan, D.},
  \bibinfo{author}{Torrisi, M.}, \bibinfo{author}{Trovato, L.},
  \bibinfo{author}{Veratelli, M.G.}, \bibinfo{author}{Wagner, S.},
  \bibinfo{year}{2015}.
\newblock \bibinfo{title}{{Big Data Analytics for Earth Sciences: the
  EarthServer approach}}.
\newblock \bibinfo{journal}{International Journal of Digital Earth}
  \bibinfo{volume}{8947}, \bibinfo{pages}{1--27}.
\newblock \URLprefix
  \url{http://www.tandfonline.com/doi/abs/10.1080/17538947.2014.1003106},
  \DOIprefix\doi{10.1080/17538947.2014.1003106}.
%Type = Article
\bibitem[{Bibring et~al.(2006)Bibring, Langevin, Mustard, Poulet, Arvidson,
  Gendrin, Gondet, Mangold, Pinet, Forget, Berth{\'{e}}, Bibring, Gendrin,
  Gomez, Gondet, Jouglet, Poulet, Soufflot, Vincendon, Combes, Drossart,
  Encrenaz, Fouchet, Merchiorri, Belluci, Altieri, Formisano, Capaccioni,
  Cerroni, Coradini, Fonti, Korablev, Kottsov, Ignatiev, Moroz, Titov, Zasova,
  Loiseau, Mangold, Pinet, Dout{\'{e}}, Schmitt, Sotin, Hauber, Hoffmann,
  Jaumann, Keller, Arvidson, Mustard, Duxbury, Forget and Neukum}]{Bibring2006}
\bibinfo{author}{Bibring, J.P.}, \bibinfo{author}{Langevin, Y.},
  \bibinfo{author}{Mustard, J.F.}, \bibinfo{author}{Poulet, F.},
  \bibinfo{author}{Arvidson, R.}, \bibinfo{author}{Gendrin, A.},
  \bibinfo{author}{Gondet, B.}, \bibinfo{author}{Mangold, N.},
  \bibinfo{author}{Pinet, P.}, \bibinfo{author}{Forget, F.},
  \bibinfo{author}{Berth{\'{e}}, M.}, \bibinfo{author}{Bibring, J.P.},
  \bibinfo{author}{Gendrin, A.}, \bibinfo{author}{Gomez, C.},
  \bibinfo{author}{Gondet, B.}, \bibinfo{author}{Jouglet, D.},
  \bibinfo{author}{Poulet, F.}, \bibinfo{author}{Soufflot, A.},
  \bibinfo{author}{Vincendon, M.}, \bibinfo{author}{Combes, M.},
  \bibinfo{author}{Drossart, P.}, \bibinfo{author}{Encrenaz, T.},
  \bibinfo{author}{Fouchet, T.}, \bibinfo{author}{Merchiorri, R.},
  \bibinfo{author}{Belluci, G.}, \bibinfo{author}{Altieri, F.},
  \bibinfo{author}{Formisano, V.}, \bibinfo{author}{Capaccioni, F.},
  \bibinfo{author}{Cerroni, P.}, \bibinfo{author}{Coradini, A.},
  \bibinfo{author}{Fonti, S.}, \bibinfo{author}{Korablev, O.},
  \bibinfo{author}{Kottsov, V.}, \bibinfo{author}{Ignatiev, N.},
  \bibinfo{author}{Moroz, V.}, \bibinfo{author}{Titov, D.},
  \bibinfo{author}{Zasova, L.}, \bibinfo{author}{Loiseau, D.},
  \bibinfo{author}{Mangold, N.}, \bibinfo{author}{Pinet, P.},
  \bibinfo{author}{Dout{\'{e}}, S.}, \bibinfo{author}{Schmitt, B.},
  \bibinfo{author}{Sotin, C.}, \bibinfo{author}{Hauber, E.},
  \bibinfo{author}{Hoffmann, H.}, \bibinfo{author}{Jaumann, R.},
  \bibinfo{author}{Keller, U.}, \bibinfo{author}{Arvidson, R.},
  \bibinfo{author}{Mustard, J.F.}, \bibinfo{author}{Duxbury, T.},
  \bibinfo{author}{Forget, F.}, \bibinfo{author}{Neukum, G.},
  \bibinfo{year}{2006}.
\newblock \bibinfo{title}{{Global Mineralogical and Aqueous Mars History
  Derived from OMEGA/Mars Express Data}}.
\newblock \bibinfo{journal}{Science} \bibinfo{volume}{312},
  \bibinfo{pages}{400--404}.
\newblock \URLprefix
  \url{http://science.sciencemag.org/content/312/5772/400.abstract},
  \DOIprefix\doi{10.1126/science.1122659}.
%Type = Article
\bibitem[{Bibring et~al.(2004)Bibring, Soufflot, Berth?, Langevin, Gondet,
  Drossart, Bouy?, Combes, Puget, Semery, Bellucci, Formisano, Moroz, Kottsov,
  Bonello, Erard, Forni, Gendrin, Manaud, Poulet, Poulleau, Encrenaz, Fouchet,
  Melchiori, Altieri, Ignatiev, Titov, Zasova, Coradini, Capacionni, Cerroni,
  Fonti, Mangold, Pinet, Schmitt, Sotin, Hauber, Hoffmann, Jaumann, Keller,
  Arvidson, Mustard and Forget}]{Bibring2004a}
\bibinfo{author}{Bibring, J.P.}, \bibinfo{author}{Soufflot, A.},
  \bibinfo{author}{Berth?, M.}, \bibinfo{author}{Langevin, Y.},
  \bibinfo{author}{Gondet, B.}, \bibinfo{author}{Drossart, P.},
  \bibinfo{author}{Bouy?, M.}, \bibinfo{author}{Combes, M.},
  \bibinfo{author}{Puget, P.}, \bibinfo{author}{Semery, A.},
  \bibinfo{author}{Bellucci, G.}, \bibinfo{author}{Formisano, V.},
  \bibinfo{author}{Moroz, V.}, \bibinfo{author}{Kottsov, V.},
  \bibinfo{author}{Bonello, G.}, \bibinfo{author}{Erard, S.},
  \bibinfo{author}{Forni, O.}, \bibinfo{author}{Gendrin, A.},
  \bibinfo{author}{Manaud, N.}, \bibinfo{author}{Poulet, F.},
  \bibinfo{author}{Poulleau, G.}, \bibinfo{author}{Encrenaz, T.},
  \bibinfo{author}{Fouchet, T.}, \bibinfo{author}{Melchiori, R.},
  \bibinfo{author}{Altieri, F.}, \bibinfo{author}{Ignatiev, N.},
  \bibinfo{author}{Titov, D.}, \bibinfo{author}{Zasova, L.},
  \bibinfo{author}{Coradini, A.}, \bibinfo{author}{Capacionni, F.},
  \bibinfo{author}{Cerroni, P.}, \bibinfo{author}{Fonti, S.},
  \bibinfo{author}{Mangold, N.}, \bibinfo{author}{Pinet, P.},
  \bibinfo{author}{Schmitt, B.}, \bibinfo{author}{Sotin, C.},
  \bibinfo{author}{Hauber, E.}, \bibinfo{author}{Hoffmann, H.},
  \bibinfo{author}{Jaumann, R.}, \bibinfo{author}{Keller, U.},
  \bibinfo{author}{Arvidson, R.}, \bibinfo{author}{Mustard, J.},
  \bibinfo{author}{Forget, F.}, \bibinfo{year}{2004}.
\newblock \bibinfo{title}{{OMEGA: Observatoire pour la min?ralogie, l'eau, les
  glaces et l'activit?}}
\newblock \bibinfo{journal}{European Space Agency, (Special Publication) ESA
  SP} \bibinfo{volume}{1240}, \bibinfo{pages}{37----49}.
%Type = Unpublished
\bibitem[{Brown et~al.(2010)Brown, Hook, Baldridge, Crowley, Bridges, Thomson,
  Marion, {de Souza Filho} and Bishop}]{Brown2010a}
\bibinfo{author}{Brown, A.J.}, \bibinfo{author}{Hook, S.J.},
  \bibinfo{author}{Baldridge, A.M.}, \bibinfo{author}{Crowley, J.K.},
  \bibinfo{author}{Bridges, N.T.}, \bibinfo{author}{Thomson, B.J.},
  \bibinfo{author}{Marion, G.M.}, \bibinfo{author}{{de Souza Filho}, C.R.},
  \bibinfo{author}{Bishop, J.L.}, \bibinfo{year}{2010}.
\newblock \bibinfo{title}{{Hydrothermal formation of Clay-Carbonate alteration
  assemblages in the Nili Fossae region of Mars}}.
\newblock \DOIprefix\doi{10.1016/j.epsl.2010.06.018},
  \href{http://arxiv.org/abs/1402.1150}{\tt arXiv:1402.1150}.
%Type = Article
\bibitem[{Carter et~al.(2013a)Carter, Poulet, Bibring, Mangold and
  Murchie}]{Carter2013}
\bibinfo{author}{Carter, J.}, \bibinfo{author}{Poulet, F.},
  \bibinfo{author}{Bibring, J.P.}, \bibinfo{author}{Mangold, N.},
  \bibinfo{author}{Murchie, S.}, \bibinfo{year}{2013}a.
\newblock \bibinfo{title}{{Hydrous minerals on Mars as seen by the CRISM and
  OMEGA imaging spectrometers: Updated global view}}.
\newblock \bibinfo{journal}{Journal of Geophysical Research E: Planets}
  \bibinfo{volume}{118}, \bibinfo{pages}{831--858}.
\newblock \URLprefix \url{http://doi.wiley.com/10.1029/2012JE004145},
  \DOIprefix\doi{10.1029/2012JE004145}.
%Type = Article
\bibitem[{Carter et~al.(2013b)Carter, Poulet, Murchie and
  Bibring}]{Carter2013a}
\bibinfo{author}{Carter, J.}, \bibinfo{author}{Poulet, F.},
  \bibinfo{author}{Murchie, S.}, \bibinfo{author}{Bibring, J.},
  \bibinfo{year}{2013}b.
\newblock \bibinfo{title}{{Automated processing of planetary hyperspectral
  datasets for the extraction of weak mineral signatures and applications to
  CRISM observations of hydrated silicates on Mars}}.
\newblock \bibinfo{journal}{Planetary and Space Science} \bibinfo{volume}{76},
  \bibinfo{pages}{53--67}.
\newblock \URLprefix
  \url{http://www.sciencedirect.com/science/article/pii/S0032063312003625},
  \DOIprefix\doi{10.1016/j.pss.2012.11.007}.
%Type = Article
\bibitem[{Chiwome(2014)}]{Oosthoek2014a}
\bibinfo{author}{Chiwome, V.}, \bibinfo{year}{2014}.
\newblock \bibinfo{title}{{Webclient-neo: Planetserver/Earthserver project
  end.}}
\newblock \bibinfo{journal}{Zenodo} \DOIprefix\doi{10.5281/zenodo.11698}.
%Type = Article
\bibitem[{Christensen et~al.(2009)Christensen, Engle, Anwar, Dickenshied, Noss,
  Gorelick and Weiss-Malik}]{Christensen2009}
\bibinfo{author}{Christensen, P.R.}, \bibinfo{author}{Engle, E.},
  \bibinfo{author}{Anwar, S.}, \bibinfo{author}{Dickenshied, S.},
  \bibinfo{author}{Noss, D.}, \bibinfo{author}{Gorelick, N.},
  \bibinfo{author}{Weiss-Malik, M.}, \bibinfo{year}{2009}.
\newblock \bibinfo{title}{{JMARS - A Planetary GIS}}.
\newblock \bibinfo{journal}{American Geophysical Union, Fall Meeting 2009,
  abstract {\#}IN22A-06} \URLprefix
  \url{http://adsabs.harvard.edu/abs/2009AGUFMIN22A..06C}.
%Type = Misc
\bibitem[{Clark et~al.(2007)Clark, Swayze, Wise, Livo, Hoefen, Kokaly and
  Sutley}]{Clark2007}
\bibinfo{author}{Clark, R.N.}, \bibinfo{author}{Swayze, G.A.},
  \bibinfo{author}{Wise, R.}, \bibinfo{author}{Livo, E.},
  \bibinfo{author}{Hoefen, T.}, \bibinfo{author}{Kokaly, R.},
  \bibinfo{author}{Sutley, S.J.}, \bibinfo{year}{2007}.
\newblock \bibinfo{title}{{USGS digital spectral library splib06a: U.S.
  Geological Survey, Digital Data Series 231}}.
\newblock \URLprefix \url{http://speclab.cr.usgs.gov/spectral.lib06.}
%Type = Article
\bibitem[{Cuadros and Michalski(2013)}]{Cuadros2013}
\bibinfo{author}{Cuadros, J.}, \bibinfo{author}{Michalski, J.R.},
  \bibinfo{year}{2013}.
\newblock \bibinfo{title}{{Investigation of Al-rich clays on Mars: Evidence for
  kaolinite-smectite mixed-layer versus mixture of end-member phases}}.
\newblock \bibinfo{journal}{Icarus} \bibinfo{volume}{222},
  \bibinfo{pages}{296--306}.
\newblock \DOIprefix\doi{10.1016/j.icarus.2012.11.006}.
%Type = Inproceedings
\bibitem[{Day and Law(2016)}]{Day2016a}
\bibinfo{author}{Day, B.H.}, \bibinfo{author}{Law, E.S.}, \bibinfo{year}{2016}.
\newblock \bibinfo{title}{{Moon Trek: NASA's new online portal for lunar
  mapping and modeling}}, in: \bibinfo{booktitle}{Annual Meeting of the Lunar
  Exploration Analysis Group}.
\newblock \URLprefix
  \url{http://www.hou.usra.edu/meetings/leag2016/pdf/5015.pdf}.
%Type = Article
\bibitem[{Ehlmann et~al.(2009)Ehlmann, Mustard, Swayze, Clark, Bishop, Poulet,
  {Des Marais}, Roach, Milliken, Wray, Barnouin-Jha and Murchie}]{Ehlmann2009}
\bibinfo{author}{Ehlmann, B.L.}, \bibinfo{author}{Mustard, J.F.},
  \bibinfo{author}{Swayze, G.A.}, \bibinfo{author}{Clark, R.N.},
  \bibinfo{author}{Bishop, J.L.}, \bibinfo{author}{Poulet, F.},
  \bibinfo{author}{{Des Marais}, D.J.}, \bibinfo{author}{Roach, L.H.},
  \bibinfo{author}{Milliken, R.E.}, \bibinfo{author}{Wray, J.J.},
  \bibinfo{author}{Barnouin-Jha, O.}, \bibinfo{author}{Murchie, S.L.},
  \bibinfo{year}{2009}.
\newblock \bibinfo{title}{{Identification of hydrated silicate minerals on Mars
  using MRO-CRISM: Geologic context near Nili Fossae and implications for
  aqueous alteration}}.
\newblock \bibinfo{journal}{Journal of Geophysical Research E: Planets}
  \bibinfo{volume}{114}, \bibinfo{pages}{E00D08}.
\newblock \URLprefix \url{http://doi.wiley.com/10.1029/2009JE003339},
  \DOIprefix\doi{10.1029/2009JE003339}.
%Type = Inproceedings
\bibitem[{Erkeling et~al.(2016)Erkeling, Luesebrink, Hiesinger, Reiss, Heyer
  and Jaumann}]{Heyer2016}
\bibinfo{author}{Erkeling, G.}, \bibinfo{author}{Luesebrink, D.},
  \bibinfo{author}{Hiesinger, H.}, \bibinfo{author}{Reiss, D.},
  \bibinfo{author}{Heyer, T.}, \bibinfo{author}{Jaumann, R.},
  \bibinfo{year}{2016}.
\newblock \bibinfo{title}{{The Multi-Temporal Database of Planetary Image Data
  (MUTED): A database to support the identification of surface changes and
  short-lived surface processes}}, in: \bibinfo{booktitle}{Planetary and Space
  Science}, pp. \bibinfo{pages}{43--61}.
\newblock \URLprefix
  \url{http://www.hou.usra.edu/meetings/lpsc2016/pdf/1852.pdf},
  \DOIprefix\doi{10.1016/j.pss.2016.03.002}.
%Type = Inproceedings
\bibitem[{Figuera et~al.(2014)Figuera, Gl{\"{a}}ser, Oberst and {De
  Rosa}}]{MarcoFiguera2014}
\bibinfo{author}{Figuera, R.M.}, \bibinfo{author}{Gl{\"{a}}ser, P.},
  \bibinfo{author}{Oberst, J.}, \bibinfo{author}{{De Rosa}, D.},
  \bibinfo{year}{2014}.
\newblock \bibinfo{title}{{Calculation of illumination conditions at the lunar
  south pole - parallel programming approach}}, in:
  \bibinfo{booktitle}{European Planetary Science Congress 2014, EPSC Abstracts,
  Vol. 9, id. EPSC2014-476}, pp. \bibinfo{pages}{EPSC2014--476}.
\newblock \URLprefix \url{http://adsabs.harvard.edu/abs/2014EPSC....9..476F}.
%Type = Article
\bibitem[{Gaddis et~al.(1997)Gaddis, Anderson, Becker, Cook, Edwards, Eliason,
  Hare, Kieffer, Lee and Matthews}]{Gaddis1997}
\bibinfo{author}{Gaddis, L.}, \bibinfo{author}{Anderson, J.},
  \bibinfo{author}{Becker, K.}, \bibinfo{author}{Cook, D.},
  \bibinfo{author}{Edwards, K.}, \bibinfo{author}{Eliason, E.},
  \bibinfo{author}{Hare, T.}, \bibinfo{author}{Kieffer, H.},
  \bibinfo{author}{Lee, E.M.}, \bibinfo{author}{Matthews, J.},
  \bibinfo{year}{1997}.
\newblock \bibinfo{title}{{An overview of the Integrated Software for Imaging
  Spectrometers(ISIS)}}.
\newblock \bibinfo{journal}{Lunar and planetary science XXVIII}
  \bibinfo{volume}{28}, \bibinfo{pages}{1997}.
%Type = Inproceedings
\bibitem[{Gaskins(2009)}]{Hogan}
\bibinfo{author}{Gaskins, T.O.M.}, \bibinfo{year}{2009}.
\newblock \bibinfo{title}{{Spatial Information Processing : Standards-Based
  Open Source Visualization Technology}}, in: \bibinfo{booktitle}{Data
  Processing}, pp. \bibinfo{pages}{1--3}.
\newblock \URLprefix
  \url{http://www.isprs.org/proceedings/xxxviii/4-w10/papers/vcgva2009{\_}03307{\_}hogan.pdf}.
%Type = Inproceedings
\bibitem[{van Gasselt et~al.(2014)van Gasselt, Morley, Houghton, Bamford,
  Ivanov, Muller, Yershov, Sidiropoulos, Gwinner, W{\"{a}}hlisch and
  Kim}]{VanGasselt2014}
\bibinfo{author}{van Gasselt, S.}, \bibinfo{author}{Morley, J.G.},
  \bibinfo{author}{Houghton, R.}, \bibinfo{author}{Bamford, S.},
  \bibinfo{author}{Ivanov, A.}, \bibinfo{author}{Muller, J.P.},
  \bibinfo{author}{Yershov, V.}, \bibinfo{author}{Sidiropoulos, P.},
  \bibinfo{author}{Gwinner, K.}, \bibinfo{author}{W{\"{a}}hlisch, M.},
  \bibinfo{author}{Kim, J.R.}, \bibinfo{year}{2014}.
\newblock \bibinfo{title}{{The iMars WebGIS – A Central Hub for Displaying
  and Distributing Co-Registered Data of Mars}}, in:
  \bibinfo{booktitle}{European Planetary Science Congress 2014}.
\newblock \DOIprefix\doi{EPSC2014-693}.
%Type = Article
\bibitem[{Golombek et~al.(2007)Golombek, Grant, Vasavada and
  Watkins}]{Golombek2007}
\bibinfo{author}{Golombek, M.}, \bibinfo{author}{Grant, J.},
  \bibinfo{author}{Vasavada, A.R.}, \bibinfo{author}{Watkins, M.},
  \bibinfo{year}{2007}.
\newblock \bibinfo{title}{{Landing Sites Proposed for the Mars Science
  Laboratory Mission}}.
\newblock \bibinfo{journal}{38th Lunar and Planetary Science Conference, (Lunar
  and Planetary Science XXXVIII), held March 12-16, 2007 in League City, Texas.
  LPI Contribution No. 1338, p.1392} \bibinfo{volume}{38},
  \bibinfo{pages}{1392}.
\newblock \URLprefix \url{http://adsabs.harvard.edu/abs/2007LPI....38.1392G}.
%Type = Article
\bibitem[{Golombek et~al.(2003)Golombek, Grant, Parker, Kass, Crisp, Squyres,
  Haldemann, Adler, Lee, Bridges, Arvidson, Carr, Kirk, Knocke, Roncoli, Weitz,
  Schofield, Zurek, Christensen, Fergason, Anderson and Rice}]{Golombek2003}
\bibinfo{author}{Golombek, M.P.}, \bibinfo{author}{Grant, J.A.},
  \bibinfo{author}{Parker, T.J.}, \bibinfo{author}{Kass, D.M.},
  \bibinfo{author}{Crisp, J.A.}, \bibinfo{author}{Squyres, S.W.},
  \bibinfo{author}{Haldemann, A.F.C.}, \bibinfo{author}{Adler, M.},
  \bibinfo{author}{Lee, W.J.}, \bibinfo{author}{Bridges, N.T.},
  \bibinfo{author}{Arvidson, R.E.}, \bibinfo{author}{Carr, M.H.},
  \bibinfo{author}{Kirk, R.L.}, \bibinfo{author}{Knocke, P.C.},
  \bibinfo{author}{Roncoli, R.B.}, \bibinfo{author}{Weitz, C.M.},
  \bibinfo{author}{Schofield, J.T.}, \bibinfo{author}{Zurek, R.W.},
  \bibinfo{author}{Christensen, P.R.}, \bibinfo{author}{Fergason, R.L.},
  \bibinfo{author}{Anderson, F.S.}, \bibinfo{author}{Rice, J.W.},
  \bibinfo{year}{2003}.
\newblock \bibinfo{title}{{Selection of the Mars Exploration Rover landing
  sites}}.
\newblock \bibinfo{journal}{Journal of Geophysical Research: Planets}
  \bibinfo{volume}{108}.
\newblock \URLprefix \url{http://doi.wiley.com/10.1029/2003JE002074},
  \DOIprefix\doi{10.1029/2003JE002074}.
%Type = Article
\bibitem[{Green et~al.(2011)Green, Pieters, Mouroulis, Eastwood, Boardman,
  Glavich, Isaacson, Annadurai, Besse, Barr, Buratti, Cate, Chatterjee, Clark,
  Cheek, Combe, Dhingra, Essandoh, Geier, Goswami, Green, Haemmerle, Head,
  Hovland, Hyman, Klima, Koch, Kramer, Kumar, Lee, Lundeen, Malaret, McCord,
  McLaughlin, Mustard, Nettles, Petro, Plourde, Racho, Rodriquez, Runyon,
  Sellar, Smith, Sobel, Staid, Sunshine, Taylor, Thaisen, Tompkins, Tseng,
  Vane, Varanasi, White and Wilson}]{Green2011}
\bibinfo{author}{Green, R.O.}, \bibinfo{author}{Pieters, C.},
  \bibinfo{author}{Mouroulis, P.}, \bibinfo{author}{Eastwood, M.},
  \bibinfo{author}{Boardman, J.}, \bibinfo{author}{Glavich, T.},
  \bibinfo{author}{Isaacson, P.}, \bibinfo{author}{Annadurai, M.},
  \bibinfo{author}{Besse, S.}, \bibinfo{author}{Barr, D.},
  \bibinfo{author}{Buratti, B.}, \bibinfo{author}{Cate, D.},
  \bibinfo{author}{Chatterjee, A.}, \bibinfo{author}{Clark, R.},
  \bibinfo{author}{Cheek, L.}, \bibinfo{author}{Combe, J.},
  \bibinfo{author}{Dhingra, D.}, \bibinfo{author}{Essandoh, V.},
  \bibinfo{author}{Geier, S.}, \bibinfo{author}{Goswami, J.N.},
  \bibinfo{author}{Green, R.}, \bibinfo{author}{Haemmerle, V.},
  \bibinfo{author}{Head, J.}, \bibinfo{author}{Hovland, L.},
  \bibinfo{author}{Hyman, S.}, \bibinfo{author}{Klima, R.},
  \bibinfo{author}{Koch, T.}, \bibinfo{author}{Kramer, G.},
  \bibinfo{author}{Kumar, A.S.K.}, \bibinfo{author}{Lee, K.},
  \bibinfo{author}{Lundeen, S.}, \bibinfo{author}{Malaret, E.},
  \bibinfo{author}{McCord, T.}, \bibinfo{author}{McLaughlin, S.},
  \bibinfo{author}{Mustard, J.}, \bibinfo{author}{Nettles, J.},
  \bibinfo{author}{Petro, N.}, \bibinfo{author}{Plourde, K.},
  \bibinfo{author}{Racho, C.}, \bibinfo{author}{Rodriquez, J.},
  \bibinfo{author}{Runyon, C.}, \bibinfo{author}{Sellar, G.},
  \bibinfo{author}{Smith, C.}, \bibinfo{author}{Sobel, H.},
  \bibinfo{author}{Staid, M.}, \bibinfo{author}{Sunshine, J.},
  \bibinfo{author}{Taylor, L.}, \bibinfo{author}{Thaisen, K.},
  \bibinfo{author}{Tompkins, S.}, \bibinfo{author}{Tseng, H.},
  \bibinfo{author}{Vane, G.}, \bibinfo{author}{Varanasi, P.},
  \bibinfo{author}{White, M.}, \bibinfo{author}{Wilson, D.},
  \bibinfo{year}{2011}.
\newblock \bibinfo{title}{{The Moon Mineralogy Mapper (M3) imaging spectrometer
  for lunar science: Instrument description, calibration, on-orbit
  measurements, science data calibration and on-orbit validation}}.
\newblock \bibinfo{journal}{Journal of Geophysical Research E: Planets}
  \bibinfo{volume}{116}, \bibinfo{pages}{E00G19}.
\newblock \URLprefix \url{http://doi.wiley.com/10.1029/2011JE003797},
  \DOIprefix\doi{10.1029/2011JE003797}.
%Type = Inproceedings
\bibitem[{Hare et~al.(2014)Hare, Keszthelyi, Gaddis and Kirk}]{Hare}
\bibinfo{author}{Hare, T.M.}, \bibinfo{author}{Keszthelyi, L.},
  \bibinfo{author}{Gaddis, L.}, \bibinfo{author}{Kirk, R.L.},
  \bibinfo{year}{2014}.
\newblock \bibinfo{title}{{Online Planetary Data and Services at USGS
  Astrogeology}}, in: \bibinfo{booktitle}{Lunar and Planetary Science
  Conference ,Vol. 45, p. 2487}.
\newblock \URLprefix
  \url{http://www.dlr.de/pf/Portaldata/6/Resources/dokumente/isprs{\_}2014/Hare{\_}MTSTC4-2014-135.pdf}.
%Type = Article
\bibitem[{Jaumann et~al.(2015)Jaumann, Tirsch, Hauber, Ansan, {Di Achille},
  Erkeling, Fueten, Head, Kleinhans, Mangold, Michael, Neukum, Pacifici, Platz,
  Pondrelli, Raack, Reiss, Williams, Adeli, Baratoux, de~Villiers, Foing,
  Gupta, Gwinner, Hiesinger, Hoffmann, Deit, Marinangeli, Matz, Mertens,
  Muller, Pasckert, Roatsch, Rossi, Scholten, Sowe, Voigt and
  Warner}]{Jaumann2015}
\bibinfo{author}{Jaumann, R.}, \bibinfo{author}{Tirsch, D.},
  \bibinfo{author}{Hauber, E.}, \bibinfo{author}{Ansan, V.},
  \bibinfo{author}{{Di Achille}, G.}, \bibinfo{author}{Erkeling, G.},
  \bibinfo{author}{Fueten, F.}, \bibinfo{author}{Head, J.},
  \bibinfo{author}{Kleinhans, M.}, \bibinfo{author}{Mangold, N.},
  \bibinfo{author}{Michael, G.}, \bibinfo{author}{Neukum, G.},
  \bibinfo{author}{Pacifici, A.}, \bibinfo{author}{Platz, T.},
  \bibinfo{author}{Pondrelli, M.}, \bibinfo{author}{Raack, J.},
  \bibinfo{author}{Reiss, D.}, \bibinfo{author}{Williams, D.},
  \bibinfo{author}{Adeli, S.}, \bibinfo{author}{Baratoux, D.},
  \bibinfo{author}{de~Villiers, G.}, \bibinfo{author}{Foing, B.},
  \bibinfo{author}{Gupta, S.}, \bibinfo{author}{Gwinner, K.},
  \bibinfo{author}{Hiesinger, H.}, \bibinfo{author}{Hoffmann, H.},
  \bibinfo{author}{Deit, L.L.}, \bibinfo{author}{Marinangeli, L.},
  \bibinfo{author}{Matz, K.D.}, \bibinfo{author}{Mertens, V.},
  \bibinfo{author}{Muller, J.}, \bibinfo{author}{Pasckert, J.},
  \bibinfo{author}{Roatsch, T.}, \bibinfo{author}{Rossi, A.},
  \bibinfo{author}{Scholten, F.}, \bibinfo{author}{Sowe, M.},
  \bibinfo{author}{Voigt, J.}, \bibinfo{author}{Warner, N.},
  \bibinfo{year}{2015}.
\newblock \bibinfo{title}{{Quantifying geological processes on Mars—Results
  of the high resolution stereo camera (HRSC) on Mars express}}.
\newblock \bibinfo{journal}{Planetary and Space Science} \bibinfo{volume}{112},
  \bibinfo{pages}{53--97}.
\newblock \URLprefix
  \url{http://linkinghub.elsevier.com/retrieve/pii/S0032063315000392},
  \DOIprefix\doi{10.1016/j.pss.2014.11.029}.
%Type = Article
\bibitem[{Liu et~al.(2016)Liu, Glotch, Scudder, Kraner, Condus, Arvidson,
  Guinness, Wolff and Smith}]{Liu2016}
\bibinfo{author}{Liu, Y.}, \bibinfo{author}{Glotch, T.D.},
  \bibinfo{author}{Scudder, N.A.}, \bibinfo{author}{Kraner, M.L.},
  \bibinfo{author}{Condus, T.}, \bibinfo{author}{Arvidson, R.E.},
  \bibinfo{author}{Guinness, E.A.}, \bibinfo{author}{Wolff, M.J.},
  \bibinfo{author}{Smith, M.D.}, \bibinfo{year}{2016}.
\newblock \bibinfo{title}{{End-member Identification and Spectral Mixture
  Analysis of CRISM Hyperspectral Data: A Case Study on Southwest Melas Chasma,
  Mars}}.
\newblock \bibinfo{journal}{Journal of Geophysical Research: Planets} ,
  \bibinfo{pages}{1--33}\URLprefix
  \url{http://doi.wiley.com/10.1002/2016JE005028},
  \DOIprefix\doi{10.1002/2016JE005028}.
%Type = Inproceedings
\bibitem[{{Lozac 'h} et~al.(2015){Lozac 'h}, Quantin-Nataf, Allemand, Bultel,
  Clenet, Harrisson, Loizeau, Ody and Thollot}]{Lozach2015}
\bibinfo{author}{{Lozac 'h}, L.}, \bibinfo{author}{Quantin-Nataf, C.},
  \bibinfo{author}{Allemand, P.}, \bibinfo{author}{Bultel, B.},
  \bibinfo{author}{Clenet, H.}, \bibinfo{author}{Harrisson, S.},
  \bibinfo{author}{Loizeau, D.}, \bibinfo{author}{Ody, A.},
  \bibinfo{author}{Thollot, P.}, \bibinfo{year}{2015}.
\newblock \bibinfo{title}{{MARSSI: a distributed information system for
  managing data of the surface of mars}}, in: \bibinfo{booktitle}{Second
  Planetary Data Workshop}.
\newblock \URLprefix
  \url{http://www.hou.usra.edu/meetings/planetdata2015/pdf/7006.pdf}.
%Type = Article
\bibitem[{{Marco Figuera}(2016)}]{MarcoFiguera2016a}
\bibinfo{author}{{Marco Figuera}, R.}, \bibinfo{year}{2016}.
\newblock \bibinfo{title}{{PlanetServer web client}}
  \DOIprefix\doi{10.5281/zenodo.200371}.
%Type = Inproceedings
\bibitem[{{Marco Figuera} et~al.(2016){Marco Figuera}, Flahaut, Gl{\"{a}}ser,
  Williams and Rossi}]{MarcoFiguera2016}
\bibinfo{author}{{Marco Figuera}, R.}, \bibinfo{author}{Flahaut, J.},
  \bibinfo{author}{Gl{\"{a}}ser, P.}, \bibinfo{author}{Williams, P.},
  \bibinfo{author}{Rossi, A.P.}, \bibinfo{year}{2016}.
\newblock \bibinfo{title}{{Water ice characterization near candidate landing
  sites at the lunar south pole}}, in: \bibinfo{booktitle}{European Lunar
  Symposium}.
\newblock \URLprefix
  \url{http://els2016.arc.nasa.gov/downloads/ELS{\_}2016{\_}Abstract{\_}Booklet{\_}v2.pdf}.
%Type = Inproceedings
\bibitem[{McGuire et~al.(2013)McGuire, Arvidson, Bishop, Brown, Cull, Green,
  Gross, Hash, Hauber, Humm, Jaumann, {Le Deit}, Malaret, Martin, Marzo,
  Morgan, Murchie, Mustard, Neukum, Parente, Platz, Roush, Seelos, Smith, Sowe,
  Tirsch, Walter, Wendt, Wiseman and Wolff}]{McGuire2013}
\bibinfo{author}{McGuire, P.}, \bibinfo{author}{Arvidson, R.},
  \bibinfo{author}{Bishop, J.}, \bibinfo{author}{Brown, A.},
  \bibinfo{author}{Cull, S.}, \bibinfo{author}{Green, R.},
  \bibinfo{author}{Gross, C.}, \bibinfo{author}{Hash, C.},
  \bibinfo{author}{Hauber, E.}, \bibinfo{author}{Humm, D.},
  \bibinfo{author}{Jaumann, R.}, \bibinfo{author}{{Le Deit}, L.},
  \bibinfo{author}{Malaret, E.}, \bibinfo{author}{Martin, T.},
  \bibinfo{author}{Marzo, G.}, \bibinfo{author}{Morgan, M.},
  \bibinfo{author}{Murchie, S.}, \bibinfo{author}{Mustard, J.},
  \bibinfo{author}{Neukum, G.}, \bibinfo{author}{Parente, M.},
  \bibinfo{author}{Platz, T.}, \bibinfo{author}{Roush, T.},
  \bibinfo{author}{Seelos, F.}, \bibinfo{author}{Smith, M.},
  \bibinfo{author}{Sowe, M.}, \bibinfo{author}{Tirsch, D.},
  \bibinfo{author}{Walter, S.}, \bibinfo{author}{Wendt, L.},
  \bibinfo{author}{Wiseman, S.}, \bibinfo{author}{Wolff, M.},
  \bibinfo{year}{2013}.
\newblock \bibinfo{title}{{Mapping Minerals on Mars with CRISM: Atmospheric and
  Photometric Correction for MRDR Map Tiles, Version 2, and Comparison to
  OMEGA}}, in: \bibinfo{booktitle}{44th Lunar and Planetary Science
  Conference}.
%Type = Article
\bibitem[{McMahon(1996)}]{McMahon1996}
\bibinfo{author}{McMahon, S.K.}, \bibinfo{year}{1996}.
\newblock \bibinfo{title}{{Overview of the Planetary Data System}}.
\newblock \bibinfo{journal}{Planetary and Space Science} \bibinfo{volume}{44},
  \bibinfo{pages}{3--12}.
\newblock \URLprefix
  \url{http://www.sciencedirect.com/science/article/pii/0032063395001018},
  \DOIprefix\doi{10.1016/0032-0633(95)00101-8}.
%Type = Article
\bibitem[{Misev et~al.(2012)Misev, Rusu and Baumann}]{Misev2012}
\bibinfo{author}{Misev, D.}, \bibinfo{author}{Rusu, M.},
  \bibinfo{author}{Baumann, P.}, \bibinfo{year}{2012}.
\newblock \bibinfo{title}{{A semantic resolver for coordinate reference
  systems}}.
\newblock \bibinfo{journal}{Lecture Notes in Computer Science (including
  subseries Lecture Notes in Artificial Intelligence and Lecture Notes in
  Bioinformatics)} \bibinfo{volume}{7236 LNCS}, \bibinfo{pages}{47--56}.
\newblock \DOIprefix\doi{10.1007/978-3-642-29247-7{\_}5}.
%Type = Article
\bibitem[{Murchie et~al.(2007)Murchie, Arvidson, Bedini, Beisser, Bibring,
  Bishop, Boldt, Cavender, Choo, Clancy, Darlington, {Des Marais}, Espiritu,
  Fort, Green, Guinness, Hayes, Hash, Heffernan, Hemmler, Heyler, Humm,
  Hutcheson, Izenberg, Lee, Lees, Lohr, Malaret, Martin, McGovern, McGuire,
  Morris, Mustard, Pelkey, Rhodes, Robinson, Roush, Schaefer, Seagrave, Seelos,
  Silverglate, Slavney, Smith, Shyong, Strohbehn, Taylor, Thompson, Tossman,
  Wirzburger and Wolff}]{Murchie2007}
\bibinfo{author}{Murchie, S.}, \bibinfo{author}{Arvidson, R.},
  \bibinfo{author}{Bedini, P.}, \bibinfo{author}{Beisser, K.},
  \bibinfo{author}{Bibring, J.P.}, \bibinfo{author}{Bishop, J.},
  \bibinfo{author}{Boldt, J.}, \bibinfo{author}{Cavender, P.},
  \bibinfo{author}{Choo, T.}, \bibinfo{author}{Clancy, R.T.},
  \bibinfo{author}{Darlington, E.H.}, \bibinfo{author}{{Des Marais}, D.},
  \bibinfo{author}{Espiritu, R.}, \bibinfo{author}{Fort, D.},
  \bibinfo{author}{Green, R.}, \bibinfo{author}{Guinness, E.},
  \bibinfo{author}{Hayes, J.}, \bibinfo{author}{Hash, C.},
  \bibinfo{author}{Heffernan, K.}, \bibinfo{author}{Hemmler, J.},
  \bibinfo{author}{Heyler, G.}, \bibinfo{author}{Humm, D.},
  \bibinfo{author}{Hutcheson, J.}, \bibinfo{author}{Izenberg, N.},
  \bibinfo{author}{Lee, R.}, \bibinfo{author}{Lees, J.}, \bibinfo{author}{Lohr,
  D.}, \bibinfo{author}{Malaret, E.}, \bibinfo{author}{Martin, T.},
  \bibinfo{author}{McGovern, J.A.}, \bibinfo{author}{McGuire, P.},
  \bibinfo{author}{Morris, R.}, \bibinfo{author}{Mustard, J.},
  \bibinfo{author}{Pelkey, S.}, \bibinfo{author}{Rhodes, E.},
  \bibinfo{author}{Robinson, M.}, \bibinfo{author}{Roush, T.},
  \bibinfo{author}{Schaefer, E.}, \bibinfo{author}{Seagrave, G.},
  \bibinfo{author}{Seelos, F.}, \bibinfo{author}{Silverglate, P.},
  \bibinfo{author}{Slavney, S.}, \bibinfo{author}{Smith, M.},
  \bibinfo{author}{Shyong, W.J.}, \bibinfo{author}{Strohbehn, K.},
  \bibinfo{author}{Taylor, H.}, \bibinfo{author}{Thompson, P.},
  \bibinfo{author}{Tossman, B.}, \bibinfo{author}{Wirzburger, M.},
  \bibinfo{author}{Wolff, M.}, \bibinfo{year}{2007}.
\newblock \bibinfo{title}{{Compact Connaissance Imaging Spectrometer for Mars
  (CRISM) on Mars Reconnaissance Orbiter (MRO)}}.
\newblock \bibinfo{journal}{Journal of Geophysical Research E: Planets}
  \bibinfo{volume}{112}, \bibinfo{pages}{E05S03}.
\newblock \URLprefix \url{http://doi.wiley.com/10.1029/2006JE002682},
  \DOIprefix\doi{10.1029/2006JE002682}.
%Type = Article
\bibitem[{Murchie et~al.(2009)Murchie, Mustard, Ehlmann, Milliken, Bishop,
  Mckeown, Dobrea, Seelos, Buczkowski, Wiseman, Arvidson, Wray, Swayze, Clark,
  Marais, Mcewen and Bibring}]{Murchie2009}
\bibinfo{author}{Murchie, S.L.}, \bibinfo{author}{Mustard, J.F.},
  \bibinfo{author}{Ehlmann, B.L.}, \bibinfo{author}{Milliken, R.E.},
  \bibinfo{author}{Bishop, J.L.}, \bibinfo{author}{Mckeown, N.K.},
  \bibinfo{author}{Dobrea, E.Z.N.}, \bibinfo{author}{Seelos, F.P.},
  \bibinfo{author}{Buczkowski, D.L.}, \bibinfo{author}{Wiseman, S.M.},
  \bibinfo{author}{Arvidson, R.E.}, \bibinfo{author}{Wray, J.J.},
  \bibinfo{author}{Swayze, G.}, \bibinfo{author}{Clark, R.N.},
  \bibinfo{author}{Marais, D.J.D.}, \bibinfo{author}{Mcewen, A.S.},
  \bibinfo{author}{Bibring, J.P.}, \bibinfo{year}{2009}.
\newblock \bibinfo{title}{{A synthesis of Martian aqueous mineralogy after 1
  Mars year of observations from the Mars Reconnaissance Orbiter}}.
\newblock \bibinfo{journal}{J. Geophys. Res} \bibinfo{volume}{114},
  \bibinfo{pages}{0--6}.
\newblock \DOIprefix\doi{10.1029/2009JE003342}.
%Type = Article
\bibitem[{Mustard et~al.(2008)Mustard, Murchie, Pelkey, Ehlmann, Milliken,
  Grant, Bibring, Poulet, Bishop, Dobrea, Roach, Seelos, Arvidson, Wiseman,
  Green, Hash, Humm, Malaret, McGovern, Seelos, Clancy, Clark, Marais,
  Izenberg, Knudson, Langevin, Martin, McGuire, Morris, Robinson, Roush, Smith,
  Swayze, Taylor, Titus and Wolff}]{Mustard2008}
\bibinfo{author}{Mustard, J.F.}, \bibinfo{author}{Murchie, S.L.},
  \bibinfo{author}{Pelkey, S.M.}, \bibinfo{author}{Ehlmann, B.L.},
  \bibinfo{author}{Milliken, R.E.}, \bibinfo{author}{Grant, J.a.},
  \bibinfo{author}{Bibring, J.P.}, \bibinfo{author}{Poulet, F.},
  \bibinfo{author}{Bishop, J.}, \bibinfo{author}{Dobrea, E.N.},
  \bibinfo{author}{Roach, L.}, \bibinfo{author}{Seelos, F.},
  \bibinfo{author}{Arvidson, R.E.}, \bibinfo{author}{Wiseman, S.},
  \bibinfo{author}{Green, R.}, \bibinfo{author}{Hash, C.},
  \bibinfo{author}{Humm, D.}, \bibinfo{author}{Malaret, E.},
  \bibinfo{author}{McGovern, J.a.}, \bibinfo{author}{Seelos, K.},
  \bibinfo{author}{Clancy, T.}, \bibinfo{author}{Clark, R.},
  \bibinfo{author}{Marais, D.D.}, \bibinfo{author}{Izenberg, N.},
  \bibinfo{author}{Knudson, a.}, \bibinfo{author}{Langevin, Y.},
  \bibinfo{author}{Martin, T.}, \bibinfo{author}{McGuire, P.},
  \bibinfo{author}{Morris, R.}, \bibinfo{author}{Robinson, M.},
  \bibinfo{author}{Roush, T.}, \bibinfo{author}{Smith, M.},
  \bibinfo{author}{Swayze, G.}, \bibinfo{author}{Taylor, H.},
  \bibinfo{author}{Titus, T.}, \bibinfo{author}{Wolff, M.},
  \bibinfo{year}{2008}.
\newblock \bibinfo{title}{{Hydrated silicate minerals on Mars observed by the
  Mars Reconnaissance Orbiter CRISM instrument.}}
\newblock \bibinfo{journal}{Nature} \bibinfo{volume}{454},
  \bibinfo{pages}{305--309}.
\newblock \URLprefix
  \url{http://www.nature.com/nature/journal/v454/n7202/pdf/nature07097.pdf},
  \DOIprefix\doi{10.1038/nature07097}.
%Type = Article
\bibitem[{Oosthoek et~al.(2015)Oosthoek, Arriazu and {Marco
  Figuera}}]{Oosthoek2015}
\bibinfo{author}{Oosthoek, J.}, \bibinfo{author}{Arriazu, P.},
  \bibinfo{author}{{Marco Figuera}, R.}, \bibinfo{year}{2015}.
\newblock \bibinfo{title}{{Shall We Send Humans to Holden Crater? How a
  Geodesic GIS Approach Can Aid the Landing Site Selection for Future Missions
  to Mars}}.
\newblock \bibinfo{journal}{First Landing Site/Exploration Zone Workshop for
  Human Missions to the Surface of Mars} \bibinfo{volume}{1879},
  \bibinfo{pages}{1049}.
\newblock \URLprefix \url{http://adsabs.harvard.edu/abs/2015LPICo1879.1049O}.
%Type = Article
\bibitem[{Oosthoek et~al.(2014)Oosthoek, Flahaut, Rossi, Baumann, Misev,
  Campalani and Unnithan}]{Oosthoek2014}
\bibinfo{author}{Oosthoek, J.}, \bibinfo{author}{Flahaut, J.},
  \bibinfo{author}{Rossi, A.}, \bibinfo{author}{Baumann, P.},
  \bibinfo{author}{Misev, D.}, \bibinfo{author}{Campalani, P.},
  \bibinfo{author}{Unnithan, V.}, \bibinfo{year}{2014}.
\newblock \bibinfo{title}{{PlanetServer: Innovative approaches for the online
  analysis of hyperspectral satellite data from Mars}}.
\newblock \bibinfo{journal}{Advances in Space Research} \bibinfo{volume}{53},
  \bibinfo{pages}{1858--1871}.
\newblock \URLprefix
  \url{http://www.sciencedirect.com/science/article/pii/S0273117713004134},
  \DOIprefix\doi{10.1016/j.asr.2013.07.002}.
%Type = Article
\bibitem[{Pelkey et~al.(2007)Pelkey, Mustard, Murchie, Clancy, Wolff, Smith,
  Milliken, Bibring, Gendrin, Poulet, Langevin and Gondet}]{Pelkey2007}
\bibinfo{author}{Pelkey, S.M.}, \bibinfo{author}{Mustard, J.F.},
  \bibinfo{author}{Murchie, S.}, \bibinfo{author}{Clancy, R.T.},
  \bibinfo{author}{Wolff, M.}, \bibinfo{author}{Smith, M.},
  \bibinfo{author}{Milliken, R.E.}, \bibinfo{author}{Bibring, J.P.},
  \bibinfo{author}{Gendrin, A.}, \bibinfo{author}{Poulet, F.},
  \bibinfo{author}{Langevin, Y.}, \bibinfo{author}{Gondet, B.},
  \bibinfo{year}{2007}.
\newblock \bibinfo{title}{{CRISM multispectral summary products: Parameterizing
  mineral diversity on Mars from reflectance}}.
\newblock \bibinfo{journal}{Journal of Geophysical Research E: Planets}
  \bibinfo{volume}{112}, \bibinfo{pages}{E08S14}.
\newblock \URLprefix \url{http://doi.wiley.com/10.1029/2006JE002831},
  \DOIprefix\doi{10.1029/2006JE002831}.
%Type = Article
\bibitem[{Pieters et~al.(2009)Pieters, Boardman, Buratti, Chatterjee, Clark,
  Glavich, Green, {Head Iii}, Isaacson, Malaret, Mccord, Mustard, Petro,
  Runyon, Staid, Sunshine, Taylor, Tompkins, Varanasi and White}]{Pieters2009}
\bibinfo{author}{Pieters, C.M.}, \bibinfo{author}{Boardman, J.},
  \bibinfo{author}{Buratti, B.}, \bibinfo{author}{Chatterjee, A.},
  \bibinfo{author}{Clark, R.}, \bibinfo{author}{Glavich, T.},
  \bibinfo{author}{Green, R.}, \bibinfo{author}{{Head Iii}, J.},
  \bibinfo{author}{Isaacson, P.}, \bibinfo{author}{Malaret, E.},
  \bibinfo{author}{Mccord, T.}, \bibinfo{author}{Mustard, J.},
  \bibinfo{author}{Petro, N.}, \bibinfo{author}{Runyon, C.},
  \bibinfo{author}{Staid, M.}, \bibinfo{author}{Sunshine, J.},
  \bibinfo{author}{Taylor, L.}, \bibinfo{author}{Tompkins, S.},
  \bibinfo{author}{Varanasi, P.}, \bibinfo{author}{White, M.},
  \bibinfo{year}{2009}.
\newblock \bibinfo{title}{{The Moon Mineralogy Mapper (M 3 ) on
  Chandrayaan-1}}.
\newblock \bibinfo{journal}{CURRENT SCIENCE} \bibinfo{volume}{96}.
%Type = Article
\bibitem[{Poulet et~al.(2005)Poulet, Bibring, Mustard, Gendrin, Mangold,
  Langevin, Arvidson, Gondet, Gomez, Berth{\'{e}}, Bibring, Langevin, Erard,
  Forni, Gendrin, Gondet, Manaud, Poulet, Poulleau, Soufflot, Combes, Drossart,
  Encrenaz, Fouchet, Melchiorri, Bellucci, Altieri, Formisano, Fonti,
  Capaccioni, Cerroni, Coradini, Korablev, Kottsov, Ignatiev, Titov, Zasova,
  Mangold, Pinet, Schmitt, Sotin, Hauber, Hoffmann, Jaumann, Keller, Arvidson,
  Mustard and Forget}]{Poulet2005}
\bibinfo{author}{Poulet, F.}, \bibinfo{author}{Bibring, J.P.},
  \bibinfo{author}{Mustard, J.F.}, \bibinfo{author}{Gendrin, A.},
  \bibinfo{author}{Mangold, N.}, \bibinfo{author}{Langevin, Y.},
  \bibinfo{author}{Arvidson, R.E.}, \bibinfo{author}{Gondet, B.},
  \bibinfo{author}{Gomez, C.}, \bibinfo{author}{Berth{\'{e}}, M.},
  \bibinfo{author}{Bibring, J.P.}, \bibinfo{author}{Langevin, Y.},
  \bibinfo{author}{Erard, S.}, \bibinfo{author}{Forni, O.},
  \bibinfo{author}{Gendrin, A.}, \bibinfo{author}{Gondet, B.},
  \bibinfo{author}{Manaud, N.}, \bibinfo{author}{Poulet, F.},
  \bibinfo{author}{Poulleau, G.}, \bibinfo{author}{Soufflot, A.},
  \bibinfo{author}{Combes, M.}, \bibinfo{author}{Drossart, P.},
  \bibinfo{author}{Encrenaz, T.}, \bibinfo{author}{Fouchet, T.},
  \bibinfo{author}{Melchiorri, R.}, \bibinfo{author}{Bellucci, G.},
  \bibinfo{author}{Altieri, F.}, \bibinfo{author}{Formisano, V.},
  \bibinfo{author}{Fonti, S.}, \bibinfo{author}{Capaccioni, F.},
  \bibinfo{author}{Cerroni, P.}, \bibinfo{author}{Coradini, A.},
  \bibinfo{author}{Korablev, O.}, \bibinfo{author}{Kottsov, V.},
  \bibinfo{author}{Ignatiev, N.}, \bibinfo{author}{Titov, D.},
  \bibinfo{author}{Zasova, L.}, \bibinfo{author}{Mangold, N.},
  \bibinfo{author}{Pinet, P.}, \bibinfo{author}{Schmitt, B.},
  \bibinfo{author}{Sotin, C.}, \bibinfo{author}{Hauber, E.},
  \bibinfo{author}{Hoffmann, H.}, \bibinfo{author}{Jaumann, R.},
  \bibinfo{author}{Keller, U.}, \bibinfo{author}{Arvidson, R.},
  \bibinfo{author}{Mustard, J.}, \bibinfo{author}{Forget, F.},
  \bibinfo{year}{2005}.
\newblock \bibinfo{title}{{Phyllosilicates on Mars and implications for early
  martian climate}}.
\newblock \bibinfo{journal}{Nature} \bibinfo{volume}{438},
  \bibinfo{pages}{623--627}.
\newblock \URLprefix \url{http://www.nature.com/doifinder/10.1038/nature04274},
  \DOIprefix\doi{10.1038/nature04274}.
%Type = Inproceedings
\bibitem[{Rossi et~al.(2016)Rossi, Hare, Baumann, Misev, Marmo, Erard, Cecconi
  and {Marco Figuera}}]{Rossi2016}
\bibinfo{author}{Rossi, P.}, \bibinfo{author}{Hare, T.},
  \bibinfo{author}{Baumann, P.}, \bibinfo{author}{Misev, D.},
  \bibinfo{author}{Marmo, C.}, \bibinfo{author}{Erard, S.},
  \bibinfo{author}{Cecconi, B.}, \bibinfo{author}{{Marco Figuera}, R.},
  \bibinfo{year}{2016}.
\newblock \bibinfo{title}{{Planetary Coordinate Reference Systems for Ogc Web
  Services. a}}, in: \bibinfo{booktitle}{47th Lunar and Planetary Science
  Conference}.
%Type = Article
\bibitem[{Spectrometer(2005)}]{Bibring2005}
\bibinfo{author}{Spectrometer, R.}, \bibinfo{year}{2005}.
\newblock \bibinfo{title}{{Mars Surface Diversity as Revealed by the}}.
\newblock \bibinfo{journal}{Science} \bibinfo{volume}{307},
  \bibinfo{pages}{1576--1581}.
\newblock \URLprefix \url{http://www.ncbi.nlm.nih.gov/pubmed/15718430},
  \DOIprefix\doi{10.1126/science.1108806}.
%Type = Article
\bibitem[{Torson and Becker(1997)}]{Torson1997}
\bibinfo{author}{Torson, J.M.}, \bibinfo{author}{Becker, K.J.},
  \bibinfo{year}{1997}.
\newblock \bibinfo{title}{{ISIS - A Software Architecture for Processing
  Planetary Images}}.
\newblock \bibinfo{journal}{Lpsc Xxviii} \bibinfo{volume}{28},
  \bibinfo{pages}{1443}.
%Type = Article
\bibitem[{Viviano-Beck et~al.(2014)Viviano-Beck, Seelos, Murchie, Kahn, Seelos,
  Taylor, Taylor, Ehlmann, Wisemann, Mustard and Morgan}]{Viviano-Beck2014}
\bibinfo{author}{Viviano-Beck, C.E.}, \bibinfo{author}{Seelos, F.P.},
  \bibinfo{author}{Murchie, S.L.}, \bibinfo{author}{Kahn, E.G.},
  \bibinfo{author}{Seelos, K.D.}, \bibinfo{author}{Taylor, H.W.},
  \bibinfo{author}{Taylor, K.}, \bibinfo{author}{Ehlmann, B.L.},
  \bibinfo{author}{Wisemann, S.M.}, \bibinfo{author}{Mustard, J.F.},
  \bibinfo{author}{Morgan, M.F.}, \bibinfo{year}{2014}.
\newblock \bibinfo{title}{{Revised CRISM spectral parameters and summary
  products based on the currently detected mineral diversity on Mars}}.
\newblock \bibinfo{journal}{Journal of Geophysical Research E: Planets}
  \bibinfo{volume}{119}, \bibinfo{pages}{1403--1431}.
\newblock \DOIprefix\doi{10.1002/2014JE004627}.

\end{thebibliography}
%\nocite{*}

\end{document}